\documentclass[aps,pra,noshowpacs,twocolumn,preprintnumbers,superscriptaddress,floatfix]{revtex4}
\pdfoutput=1

\usepackage{graphicx}
\usepackage{amsmath}
\usepackage{color}

\usepackage{algpseudocode}
\usepackage{algorithm}

\usepackage{multirow}
\usepackage{placeins}

\begin{document}

\title{A Benchmark Test of Boson Sampling on Tianhe-2 Supercomputer}

\author{Junjie Wu}
\email{junjiewu@nudt.edu.cn}
\affiliation{Institute for Quantum Information \& State Key Laboratory of High Performance Computing, College of Computer, National University of Defense Technology, Changsha 410073, China}
\author{Yong Liu}
\affiliation{Institute for Quantum Information \& State Key Laboratory of High Performance Computing, College of Computer, National University of Defense Technology, Changsha 410073, China}
\author{Baida Zhang}
\affiliation{Institute for Quantum Information \& State Key Laboratory of High Performance Computing, College of Computer, National University of Defense Technology, Changsha 410073, China}
\author{Xianmin Jin}
\email{xianmin.jin@sjtu.edu.cn}
\affiliation{State Key Laboratory of Advanced Optical Communication Systems and Networks, School of Physics and Astronomy, Shanghai Jiao Tong University, Shanghai 200240, China}
\affiliation{Synergetic Innovation Center of Quantum Information and Quantum Physics, University of Science and Technology of China, Hefei, Anhui 230026, China}
\author{Yang Wang}
\affiliation{Institute for Quantum Information \& State Key Laboratory of High Performance Computing, College of Computer, National University of Defense Technology, Changsha 410073, China}
\author{Huiquan Wang}
\affiliation{Institute for Quantum Information \& State Key Laboratory of High Performance Computing, College of Computer, National University of Defense Technology, Changsha 410073, China}
\author{Xuejun Yang}
\email{xjyang@nudt.edu.cn}
\affiliation{Institute for Quantum Information \& State Key Laboratory of High Performance Computing, College of Computer, National University of Defense Technology, Changsha 410073, China}

\begin{abstract}
  Boson sampling, thought to be intractable classically, can be solved by a quantum machine composed of merely generation, linear evolution and detection of single photons. Such an analog quantum computer for this specific problem provides a shortcut to boost the absolute computing power of quantum computers to beat classical ones. However, the capacity bound of classical computers for simulating boson sampling has not yet been identified. Here we simulate boson sampling on the Tianhe-2 supercomputer which occupied the first place in the world ranking six times from 2013 to 2016. We computed the permanent of the largest matrix using up to 312,000 CPU cores of Tianhe-2, and inferred from the current most efficient permanent-computing algorithms that an upper bound on the performance of Tianhe-2 is one 50-photon sample per $\sim$100 min. In addition, we found a precision issue with one of two permanent-computing algorithms.
\end{abstract}

\keywords{Boson sampling, permanent, quantum supremacy, quantum computing, supercomputer}

\maketitle

\section{Introduction}

Universal quantum computers promise to substantially outperform classical computers~\cite{Deutsch1997,Nielsen2011}. However, building them has been experiencing challenges in practices, due to the stringent requirements of high-fidelity quantum gates and scalability. For example, Shor's algorithm~\cite{Shor1994}, which solves the integer factorization problem, is one of the most attractive quantum algorithms because of its potential to crack current mainstream RSA cryptosystems. The key size crackable on classical computers is 768 bits~\cite{Kleinjung2010}. This size, however, requires millions of qubits for a quantum computer to do the factorization~\cite{Fowler2012}, far from current technology~\cite{Martin2012,Lanyon2007,Politi2009,Parker2000,Lu2007,Monz2016,Vandersypen2001}. This gap motivates research into purpose-specific quantum computation with quantum speedup and more favorable experimental conditions.

Boson sampling~\cite{Aaronson2011} is a specific quantum computation thought to be an outstanding candidate for beating the most powerful classical computer in the near term. It samples the distribution of bosons output from a complex interference network. Unlike universal quantum computation, quantum boson-sampling seems to be more straightforward, since it only requires identical bosons, linear evolution and measurement. As for classical computers, the distribution can be obtained by computing permanents of matrices derived from the unitary transformation matrix of the network~\cite{Scheel2008}, in which the most time-consuming task for the simulation of boson sampling is the calculation of permanents. However, computing the permanent has been proved as a \#P-hard task on classical computers~\cite{Aaronson2011}. This motivates massive advances in building larger quantum boson-sampling machines to outperform classical computers, including principle-of-proof experiments~\cite{Spring2013,Broome2013,Tillmann2013,Crespi2013}, simplified models that are easy to implement~\cite{Aaronson2013blog,Lund2014,Bentivegnae2015}, implementation techniques~\cite{Motes2014,Spring2016,Carolan2015,Defienne2016}, robustness of boson sampling~\cite{Rohde2012err,Rohde2012,Rahimi2016,Motes2013,Motes2015}, validation of large scale boson sampling~\cite{Aaronson2014,Spagnolo2014,Wang2016}, varied models for other quantum states~\cite{Rohde2015,Seshadreesan2015,Olson2015}, etc.

\begin{figure*}
  \centering
  \includegraphics[width=.85\linewidth]{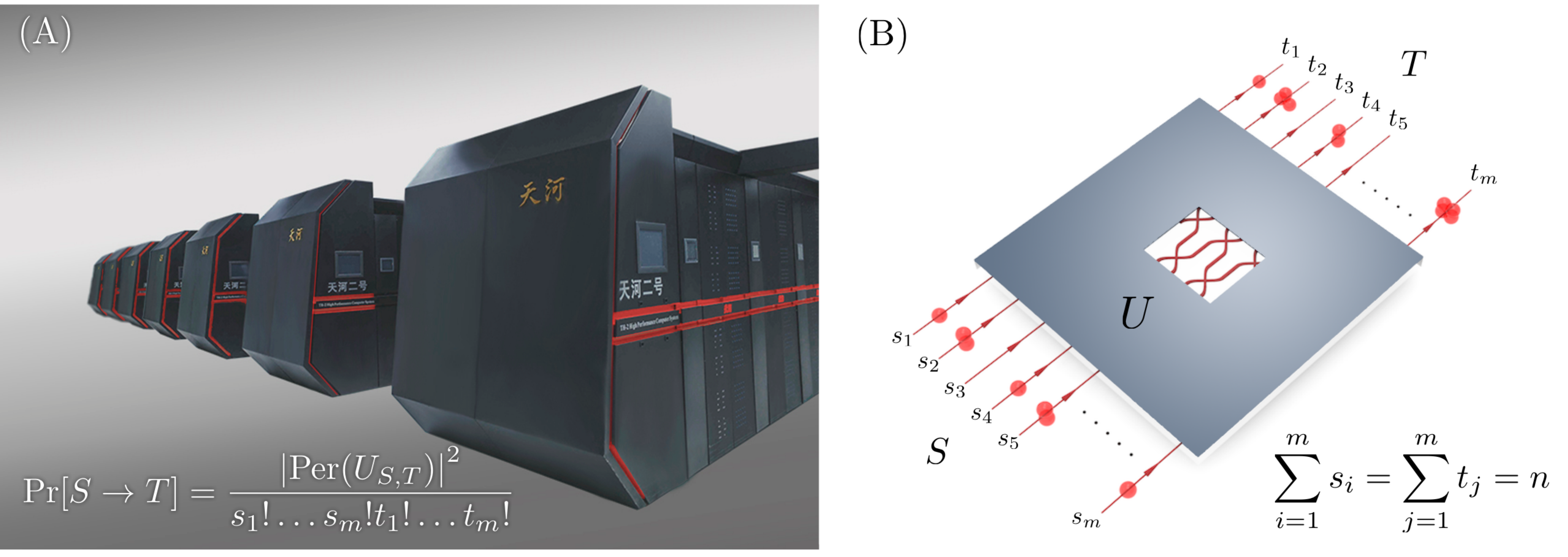}
  \caption{\footnotesize{\textbf{A schematic view of computational task with the Tianhe-2 supercomputer (A) and a quantum boson-sampling machine (B).}
      A quantum boson-sampling machine obtains an $n$-photon sample $T$ directly through a measurement on the $m$ output ports from the network that described by a unitary matrix $U$ with input $S$. To simulate the generation of a sample on Tianhe-2, it is necessary to compute the probability $\text{Pr}[S\rightarrow T]$, in which the main time-consuming task is to calculate the permanent of an $n\times n$ sub-matrix $U_{S,T}$ of $U$. The capacity of computing the permanent on Tianhe-2 is employed to benchmark the state of the art and set an upper bound on the classical execution time to be beaten by quantum boson-sampling machine.}}
  \label{fig:Tianhe2AndBosonSampling}
\end{figure*}

However, what is the capacity bound of a state-of-the-art classical computer for simulating boson sampling? This bound indicates the condition on which a quantum boson-sampling machine will surpass classical computers.

Given an $m\times m$ unitary matrix $U$ and $n$ indistinguishable bosons (as shown in FIG.~\ref{fig:Tianhe2AndBosonSampling}(B)), the simulation on classical computers is to generate samples from the output distribution described by Equation~(\ref{eq:bosonSamplingOutputDistribution}).
\begin{equation}
  \label{eq:bosonSamplingOutputDistribution}
  \text{Pr}[S\rightarrow T]=\frac{\left|\text{Per}(U_{S,T})\right|^2}{s_1!\ldots s_m!t_1!\ldots t_m!}
\end{equation}
where $S=|s_1,\ldots, s_m\rangle$ is a given input state with $s_i$ bosons in the $i^\text{th}$ input port, $T=|t_1,\ldots, t_m\rangle$ is an output state with $t_j$ bosons in the $j^\text{th}$ output port, and $U_{S,T}$ is an $n\times n$ sub-matrix derived from $U$~\cite{Aaronson2011}. The permanent calculation is the most time-consuming task in the simulation of boson sampling on a classical computer, because it is the source of the hardness in the complexity conjecture~\cite{Aaronson2011}. Therefore, the performance of computing the permanent of the $n\times n$ sub-matrix, $U_{S,T}$, is an upper bound on the performance of generating an $n$-photon sample from the distribution $\text{Pr}[S\rightarrow T]$. In this paper, we evaluate this upper bound by testing two most efficient permanent-computing algorithms on the Tianhe-2 supercomputer~\cite{Liao2014} (FIG.~\ref{fig:Tianhe2AndBosonSampling}(A)). The results show that Tianhe-2 requires about 100 min to generate one 50-photon sample.

\section{Speed Performance}

The two most efficient permanent-computing algorithms, Ryser's algorithm and BB/FG's algorithm, are both in the time complexity of $O(n^2\cdot 2^n)$.

We implemented Ryser's algorithm and BB/FG's algorithm (see the supplementary material for details), and ran them on the Tianhe-2 supercomputer. This supercomputer consists of 16,000 computing nodes, each containing three CPUs and two co-processors, denoted as MIC. The programs were tested under two types of configurations: running with only CPUs, or hybrid running with both CPUs and MICs.

We ran Ryser's algorithms with the number of nodes ranging from 2,048 to 13,000, as shown in TABLE~\ref{tab:largeScaleTests}. It is difficult for a system of very large scale to complete long-running-time execution, because the system reliability becomes worse as the number of processing units increases~\cite{Yang2012}. Occasionally, slow nodes would prolong the total execution time. This phenomenon can be seen from the data in TABLE~\ref{tab:largeScaleTests}, since the time used is not reduced in proportion (more specifically, the time used for a $46\times 46$ permanent using 4,096, 8,192 and 13,000 nodes). Up to now, the $48\times 48$ matrix's permanent is the largest problem computed on 8,192 nodes, which accounts for more than half of the nodes of Tianhe-2, and the 13,000-node test uses 81.25\% CPUs of Tianhe-2, the largest amount of computing resources ever, for the boson-sampling problem.

\begin{table}[!htb]
  \centering
  \caption{\footnotesize{\textbf{Large-scale tests of Ryser's algorithm on the Tianhe-2 supercomputer.}
      Each row of the table gives the result of testing an $n\times n$ matrix with $C$ CPU cores of $P$ computing nodes. Items in ``Predicted time (s)'' column are intervals of predicted execution time through fitting results of tests on the subsystem. The red items show that the execution time is longer than expected, suggesting the existence of slow nodes.}}
  \label{tab:largeScaleTests}
    \begin{tabular}{ccccc}
    \hline
    \hline
    $n$ & $P$ & $C$ & Execution time (s) & Predicted time (s) \\
    \hline
    40 &      4,096 &     98,304 &  24.863563 & [22.650, 27.758]\\
    40 &      8,192 &    196,608 &  {\color{red}14.105145} & {\color{red}[11.076, 13.726]}\\
    45 &      2,048 &     49,152 &  1967.7603 & [1875.8, 2273.5]\\
    45 &      4,096 &     98,304 &  984.02218 & [917.34, 1124.2]\\
    45 &      4,096 &     98,304 &  981.25833 & [917.34, 1124.2]\\
    45 &      4,096 &     98,304 &  986.09726 & [917.34, 1124.2]\\
    45 &      4,096 &     98,304 &  981.44862 & [917.34, 1124.2]\\
    45 &     13,000 &    312,000 &  {\color{red}450.62642} & {\color{red}[278.54, 347.72]}\\
    46 &      4,096 &     98,304 &  2096.3798 & [1917.1, 2349.4]\\
    46 &      8,192 &    196,608 &  1054.9386 & [937.54, 1161.7]\\
    46 &     13,000 &    312,000 &  {\color{red}1034.0247} & {\color{red}[582.13, 726.71]}\\
    48 &      8,000 &    192,000 &  4630.0842 & [4184.5, 5183.4]\\
    48 &      8,000 &    192,000 &  4628.1068 & [4184.5, 5183.4]\\
    48 &      8,000 &    192,000 &  4657.8987 & [4184.5, 5183.4]\\
    48 &      8,000 &    192,000 &  4648.9111 & [4184.5, 5183.4]\\
    48 &      8,000 &    192,000 &  4627.6037 & [4184.5, 5183.4]\\
    48 &      8,192 &    196,608 &  4530.5931 & [4083.3, 5060.0]\\
    48 &      8,192 &    196,608 &  4498.4557 & [4083.3, 5060.0]\\
    \hline
    \hline
\end{tabular}
\end{table}

\begin{figure*}[!htb]
  \centering
  \includegraphics[width=.9\linewidth]{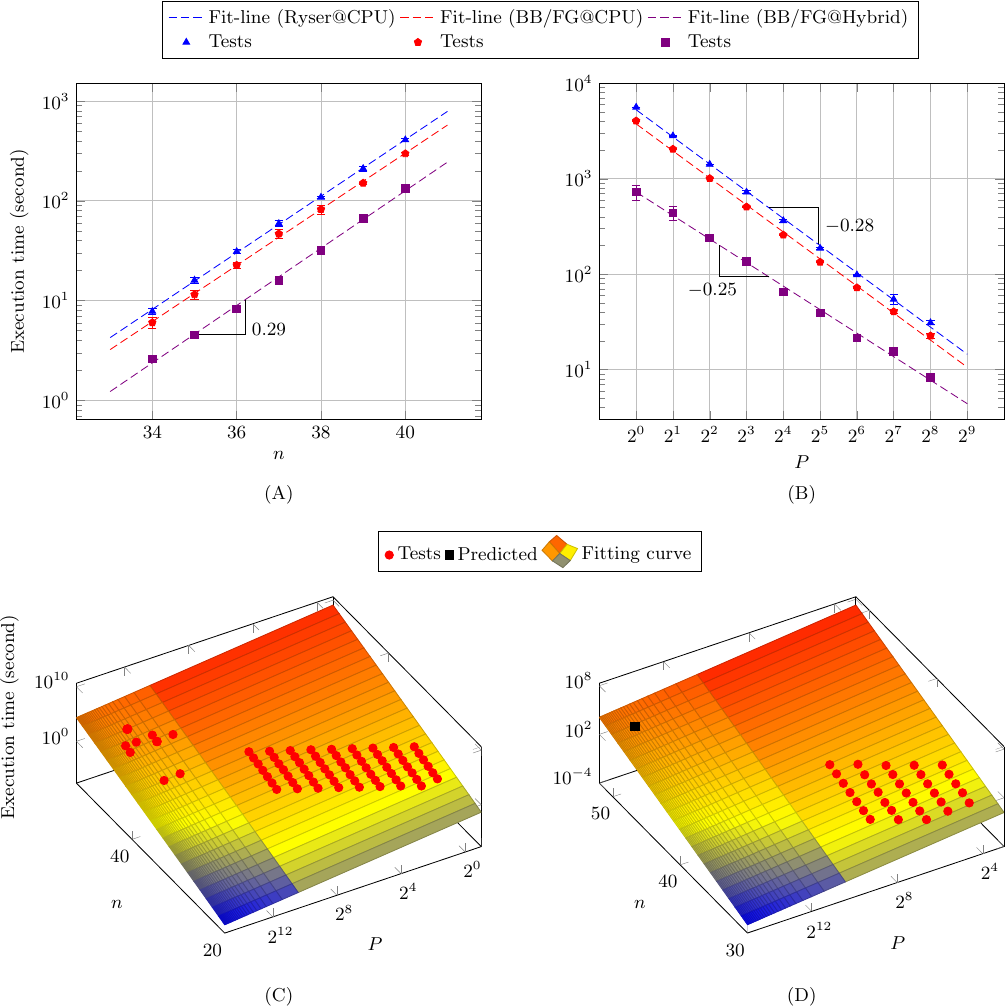}
  \caption{\footnotesize{\textbf{Scalability.}
      (A) and (B) show the execution time for $P$ nodes to compute the permanent of an $n\times n$ matrix. The ``@CPU'' terms are the results obtained from the running only on CPUs, and the ``Hybrid'' terms are the execution time on CPUs accelerated with co-processors. The slope of the fit-lines indicates the execution time increase by $\sim$$1.95$ times when $n$ increases by 1, and decrease by $0.52\sim 0.56$ times when the number of computing nodes doubles. The error bars, where visible, denote one standard deviation. (C) is the fitted execution time of Ryser's algorithm for $P$ nodes to compute the permanent of an $n\times n$ matrix. The adjusted R-square statistic is 0.9996, indicating a good fit. (More details are given in the supplementary material.) (D) is the fitted execution time of BB/FG's algorithm. The black point is the predicted time used for the full system of Tianhe-2 to compute the permanent of a $50\times 50$ matrix.}}
  \label{fig:executionTime}
\end{figure*}

Both the algorithms were tested on a 256-node subsystem of Tianhe-2 (which still has a theoretical peak performance of 848.4 teraflops, which may be ranked in top 160 in the Top500 list in November 2015) to evaluate the scalability through tests under different parameter combinations of matrix order and number of the parallel scale. As shown by the results in FIG.~\ref{fig:executionTime}, execution time increases by $\sim$1.95($\approx10^{0.29}$) times when $n$ increases by 1 (part A), and decreases with a nearly linear speedup when the number of nodes used increases (part B). These results reflect the fact that our programs are very scalable with only a little extra cost from the parts that cannot be parallelized.

To evaluate the scalability in more detail, we propose a fitting equation of the execution time involving both problem size and computing resources, as shown in Equation~(\ref{eq:fittingEquation}). 
\begin{equation}\label{eq:fittingEquation}
  T=\frac{an^22^n}{P^b}
\end{equation}
where $a$ and $b$ are the fitting coefficients, and $T$ represents the execution time of computing the permanent of an $n\times n$ matrix with $P$ computing nodes. We fitted the execution time of Ryser's algorithm with data tested on $1$ to $13,000$ nodes, as shown in FIG.~\ref{fig:executionTime}(C). The fitted $b$ is $0.9924$, showing the good scalability of our programs again. This comes from the feature of the algorithm. The amount of communication is just $O(n^2P)$, while that of computation is $O(n^22^n)$, growing much faster than that of communication.

\begin{figure*}[!htb]
  \centering
  \includegraphics[width=.9\linewidth]{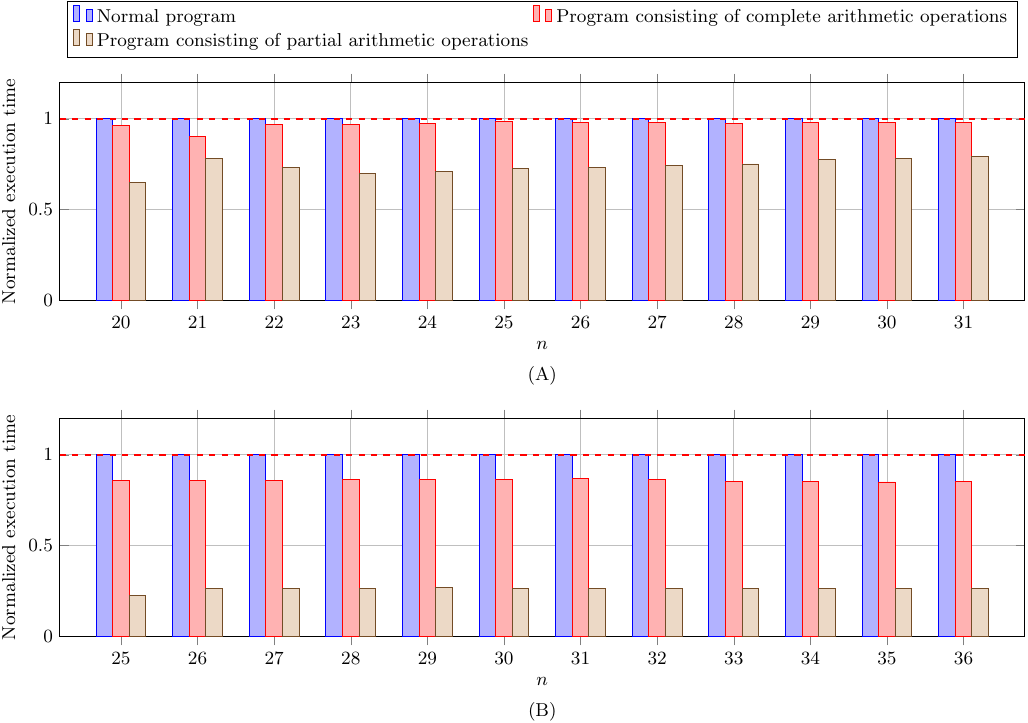}
  \caption{\footnotesize{\textbf{Efficiency for utilizing CPUs and co-processors.}
      (A) and (B) are normalized execution time on CPUs and co-processors respectively. The ratios of the execution time of the two modified programs over that of normal ones are considered as efficiency. The average efficiency for CPUs is 97.0\% and 74.0\% following the order in the figure, and that for co-processors is 86.0\% and 26.3\%.}}
  \label{fig:efficiency}
\end{figure*}

To evaluate the efficiency at which our programs utilize CPUs and co-processors, we removed implementation-related instructions from our programs, only leaving essential arithmetic operations connected to the computational complexity. The execution time of this kind of program is viewed as a baseline. FIG.~\ref{fig:efficiency} shows that our programs have exploited the performance from CPUs and co-processors in considerable efficiency. More evaluations and further analysis can be found in the supplementary material.

We used the fitted execution time of BB/FG's algorithm, not that of Ryser's due to its precision issue (discussed in the next section), to analyze the capacity bound of the full system of Tianhe-2. The fitting equation is shown in Equation~(\ref{eq:fitting}).
\begin{equation}\label{eq:fitting}
  T=\frac{9.805\cdot n^22^n}{P^{0.8782}\times10^{12}}
\end{equation}
The fitting result, in FIG.~\ref{fig:executionTime}(D), suggests that the execution time for the full system of Tianhe-2 with all CPUs and co-processors to compute the permanent of a $50 \times 50$ matrix would be about 93.8 min, while the 95\% confidence bound of this prediction is [77.41, 112.44] min, which means that an upper bound of Tianhe-2 is one 50-photon sample per $\sim$100 min.

\section{Precision Performance}

\begin{figure*}[!htb]
  \centering
  \begin{minipage}[t]{.48\linewidth}
    \includegraphics[width=.9\linewidth]{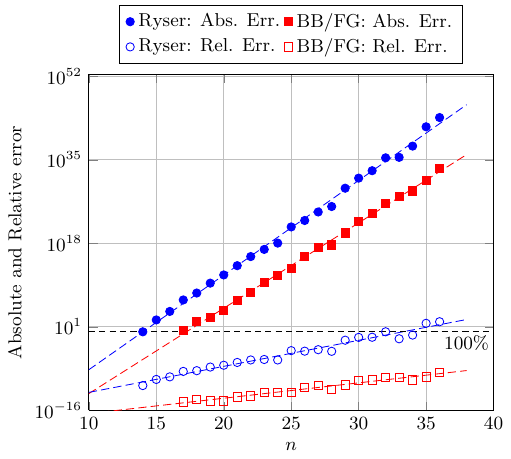}
    \centering
    \caption{\footnotesize{\textbf{Errors of computing permanents of $n\times n$ all-ones matrices.}
        An absolute error for a permanent of the $n\times n$ all-ones matrix is the substraction between the computed value and the theoretical value ($n!$). Both the errors of Ryser's algorithm and BB/FG's algorithm grow exponentially with the increase of $n$. The relative errors are computed through dividing the absolute errors by the theoretical values. Unfortunately, the relative error rate of Ryser's algorithm for a $32\times 32$ all-ones matrix reaches beyond 100\%.}}
    \label{fig:errorsAllOneMatrices}
  \end{minipage}\hfill%
  \begin{minipage}[t]{.48\linewidth}
    \centering
    \includegraphics[width=.9\linewidth]{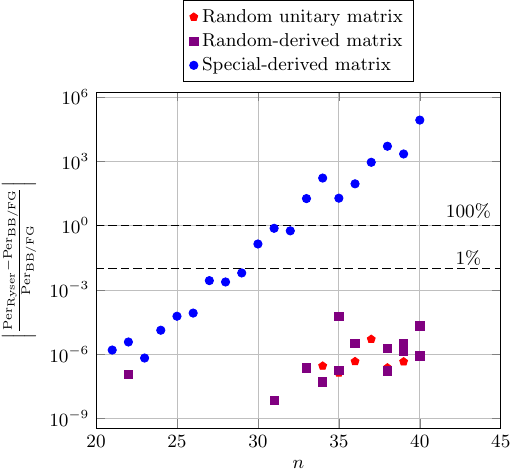}
    \caption{\footnotesize{\textbf{Errors of computing permanents of $n\times n$ random matrices.}
        ``Random unitary matrix'' denotes results of matrices corresponding to that each input (output) port of an interference network injecting (ejecting) one and only one boson. ``Random-derived matrix'' denotes results of matrices randomly derived from a $100\times 100$ unitary matrix, and ``Special-derived matrix'' denotes results of matrices specially chosen to reveal more errors. The two algorithms give very similar results for the former two matrices. However, the specially derived matrices are still overburdened by accumulated rounding errors. The relative difference between the two algorithms achieves 13.9\% for the $30\times 30$ specially derived matrix tested.
      }}
    \label{fig:errorsRandomMatrix}
  \end{minipage}%
\end{figure*}

A precision issue was found during the test of Ryser's algorithm. The limited word length of classical electricity computers leads to an accumulated rounding error (see supplementary material for details). To evaluate the errors of the two algorithms, a special type of matrix, an all-ones matrix, was used for its theoretical value is easy to obtain. The result reveals that, as shown in FIG.~\ref{fig:errorsAllOneMatrices}, the relative error of Ryser's algorithm reaches nearly 100\% when $n\geq 30$.

To confirm the precision issue, we generated three types of random matrices. The randomness of the built matrices make their permanents hard to evaluate; thus we compare the results by Ryser's and BB/FG's algorithms to verify the computation results. As shown in FIG.~\ref{fig:errorsRandomMatrix}, the errors of random unitary matrices and randomly derived matrices are marginal, but that of the specially derived matrix chosen was still very large. The growing speed of errors of specially derived matrices is exponential. Thus the precision issue may not be omitted in the future when using Ryser's algorithm for classical validation.


\section{Discussion}

In this paper, we have inferred an upper bound on the performance of simulating boson sampling of the Tianhe-2 supercomputer. Because Tianhe-2 was the fastest classical computer from 2013 to 2016, this bound was for classical computers at that time. Since the performance of classical computer is continually improving by hardware advances and software optimizations, the bound becomes higher and higher. In addition, the error evaluation of the two algorithms suggests that when using classical computers for the verification of the experiment, BB/FG's algorithm should be the first choice.

\section*{Acknowledgements}
We gratefully acknowledge the help from the National Supercomputer Center in Guangzhou. We would like to thank Scott Aaronson for his kind help. We appreciate the helpful discussion with Yunfei Du, Ping Xu, Xun Yi, Xuan Zhu, Jiangfang Ding, Hongjuan He, Yingwen Liu, Dongyang Wang and Shichuan Xue. We also thank the anonymous referee for the helpful comments. This work was supported by the National Natural Science Foundation of China (NSFC) Nos. 61632021 and 61221491, and the Open Fund from the State Key Laboratory of High Performance Computing of China (HPCL) No.201401-01.

\bibliographystyle{unsrt}

\clearpage
\onecolumngrid

\setcounter{section}{0}
\setcounter{equation}{0}
\setcounter{figure}{0}
\setcounter{table}{0}
\renewcommand{\theequation}{S\arabic{equation}}
\renewcommand{\thefigure}{S\arabic{figure}}
\renewcommand{\thetable}{S\Roman{table}}
\renewcommand{\bibnumfmt}[1]{[S#1]}
\renewcommand{\citenumfont}[1]{S#1}

\begin{center}\bf\large
  Supplementary Information: \\ A Benchmark Test of Boson Sampling on Tianhe-2 Supercomputer
\end{center}

\section{Parallelization and Optimization}
\label{app:methodology}

We implemented the two most-efficient permanent-computing algorithms, Ryser's algorithm and BB/FG's algorithm. The equation form for these two algorithms are  shown in Equation~\eqref{eq:ryser} and Equation~\eqref{eq:bbfg} respectively.
\begin{equation}\label{eq:ryser}
  \text{Per}(A)=(-1)^n\sum_{S\subseteq\{1,2,...,n\}}(-1)^{|S|}\prod_{i=1}^n\sum_{j\in S}a_{ij}
\end{equation}

\begin{equation}\label{eq:bbfg}
  \text{Per}(A)=\frac{\sum_\delta(\prod_{k=1}^n\delta_k)\prod_{i=1}^n\sum_{j=1}^n\delta_ja_{ij}}{2^{n-1}}
\end{equation}
where $A=\{a\}_{n\times n}$ are the matrix whose permanent is to be computed, $\delta=\{\delta_1, \delta_2, ..., \delta_n\}$ is the auxiliary array with $\delta_1=1$ and $\delta_i\in \{-1,1\}$ for $2\leq i\leq n$. Obviously both the algorithms are in the time complexity of $O(n^2\cdot 2^n)$. Note that matrices of Boson sampling are complex matrices where the approximation algorithm (JSV algorithm~\cite{Jerrum2004}, for example) could not be applied.

\begin{figure*}
  \centering
  \includegraphics[width=\linewidth]{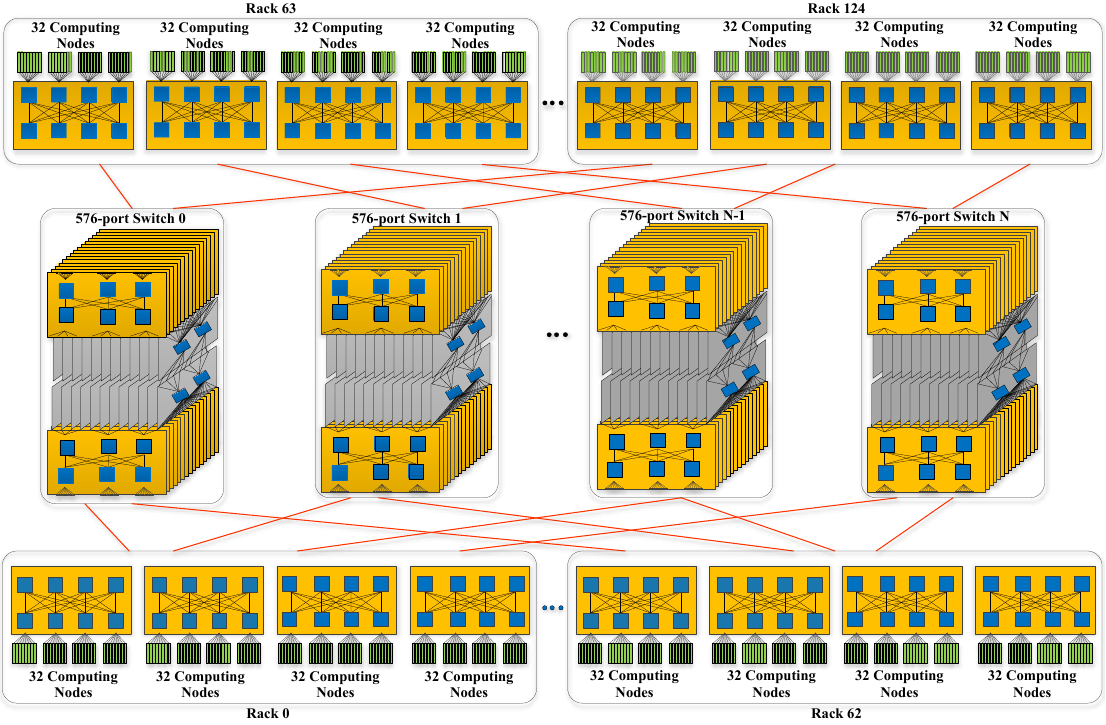}
  \caption{\footnotesize{\textbf{The architecture of Tianhe-2 supercomputer~\protect\cite{Liao2014s}.}
      Tianhe-2 has 125 computing racks, each of which has 128 computing nodes. Therefore, Tianhe-2 has $16,000$ computing nodes in total, which are connected to switches following a customized fat-tree topology.
    }}
  \label{fig:Tianhe2Architecture}
\end{figure*}

\begin{figure*}
  \centering
  \includegraphics[width=.45\linewidth]{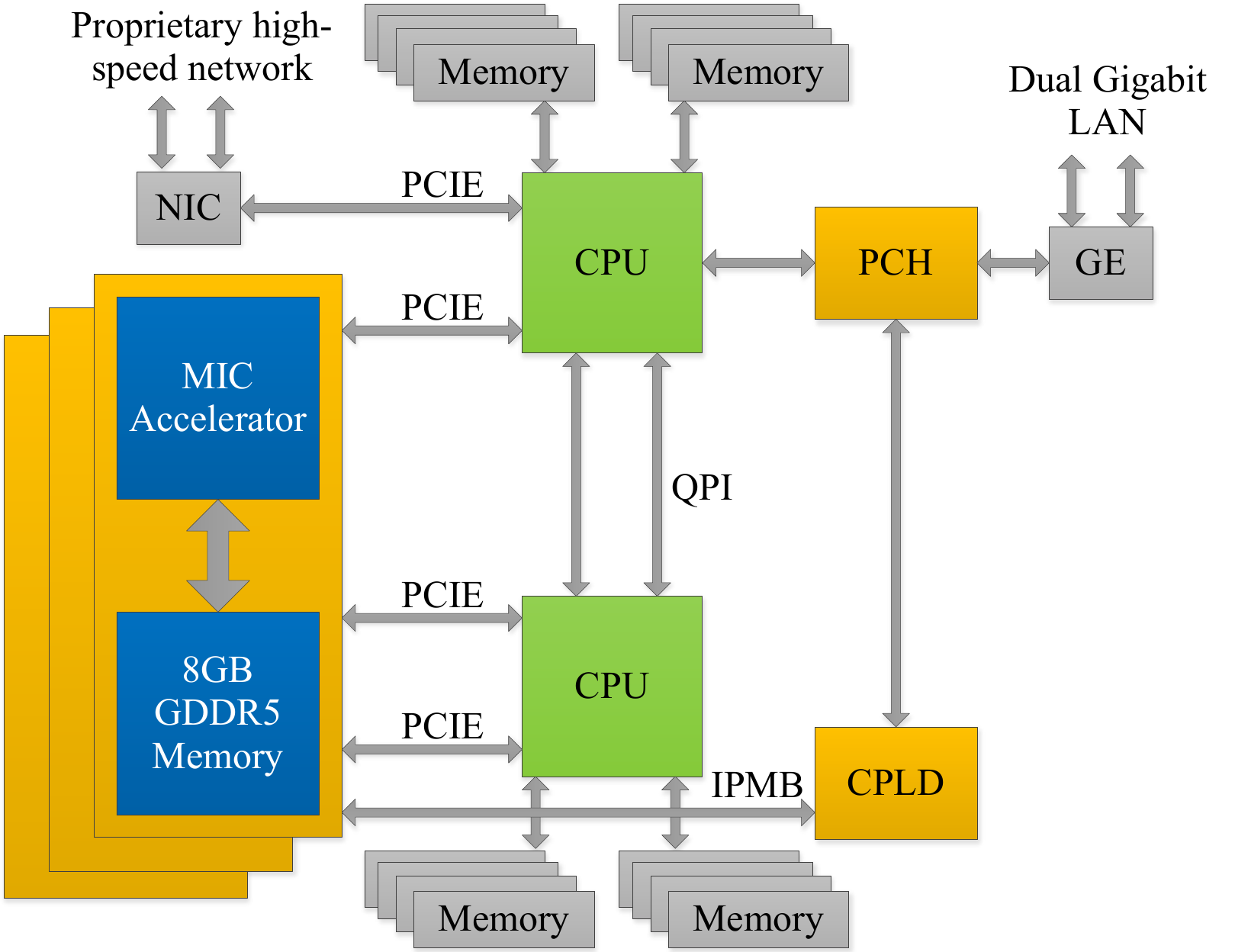}
  \caption{\footnotesize{\textbf{The architecture of a computing node in Tianhe-2 supercomputer~\protect\cite{Liao2014s}.}
      Each node comprises 2 CPUs and 3 MIC accelerators.
    }}
  \label{fig:Tianhe2Node}
\end{figure*}

To obtain the capacity bound of computing permanents classically, we tried exploiting as many computing resources of Tianhe-2 as possible. We parallelized and optimized the programs of Ryser's algorithm and BB/FG's algorithm on Tianhe-2 supercomputer~\cite{Liao2014s}. The architecture of Tianhe-2 is shown in FIG.~\ref{fig:Tianhe2Architecture} and FIG.~\ref{fig:Tianhe2Node}. The performance parameters of Tianhe-2 is listed in TABLE~\ref{tab:PerfPara}. Before the large-scale test, we use a subsystem of Tianhe-2 with no more than 256 computing nodes to optimize and evaluate our programs. Ryser's algorithm was tested with only CPUs, while BB/FG's algorithm was tested in two types of configurations: running with only CPUs, or hybrid running with both CPUs and the co-processors denoted as MIC. Note that this 256-node subsystem still has a theoretical peak performance of 848.4 Teraflops that may be ranked in top 160 in the Top500 list in Nov. 2015.

The key of gaining performance improvement from a parallel computer system is to implement the parallelism of the program and guarantee its good scalability when using massive computing resources. Both Ryser's and BB/FG's algorithms could be easily implemented with effective parallelism. Though the two algorithms being executed in Gray code order decreases the time complexity by a factor $n$ compared with the original time complexity of $O(n^2\cdot 2^n)$, this leads to so serious data dependency and memory cost that little parallelism could be implemented. Thus the algorithms with original execution order is chosen for supercomputer platform.

\subsection{Utilizing Multiple CPUs (and CPU cores)}

We used MPI library to generate multiple processes, exploiting the parallelism between CPUs, and OpenMP to produce multiple threads, exploiting that between CPU cores. As shown in Algorithm~\ref{alg:ParallelRyserProg} and Algorithm~\ref{alg:ParallelBBFGProg}, different processes have no data dependency except the data broadcast and the final reduce operation (line~\ref{algline:RyserBroadcast},~\ref{algline:RyserReceive} and~\ref{algline:RyserReduce} of Algorithm~\ref{alg:ParallelRyserProg} and Algorithm~\ref{alg:ParallelBBFGProg}) in which the main process (process 0) gathers data from other processes. This brings the good scalability of the parallel programs.

\begin{algorithm}[tbp]
  \caption{\small Parallelized Ryser's algorithm}
  \label{alg:ParallelRyserProg}
  \begin{algorithmic}[1]
    \Require $A$: an $n\times n$ matrix; $P$: number of processes
    \Ensure $Per$: Permanent of A
    \State Initialization;\Comment{For parallelization using MPI}
    \State Set $rank$ to current process ID;
    \If{$rank==0$}
    \State Generate matrix and broadcast data;\label{algline:RyserBroadcast}
    \Else
    \State Receive data from process 0;\label{algline:RyserReceive}
    \EndIf
    \State $length=\lceil (2^n-1)/P\rceil$;
    \State $start=rank*length$;
    \State $end=\min (start+length,2^n-1)$;
    \For{$iter=start$ to $end$}\Comment{Parallelized using OpenMP}
    \State Generate set $S$ ($i\in S$ iff $i^\text{th}$ bit of $iter$ is 1);
    \State $Per=Per+(-1)^{|S|}\prod_{i=1}^n\sum_{j \in S}a_{ij}$;
    \EndFor
    \State Sum $Per$s from all processes \& store result on process 0;\label{algline:RyserReduce}
    \If{$n$ is odd \textbf{and} $rank==0$}
    \State $Per=-1*Per$;
    \EndIf
    \State Process 0 outputs $Per$;
  \end{algorithmic}
\end{algorithm}

\begin{algorithm}[tbp]
  \caption{\small Parallelized BB/FG's algorithm}
  \label{alg:ParallelBBFGProg}
  \begin{algorithmic}[1]
    \Require $A$: an $n\times n$ matrix; $P$: number of processes
    \Ensure $Per$: Permanent of A
    \State Initialization;\Comment{For parallelization using MPI}
    \State Set $rank$ to current process ID;
    \If{$rank==0$}
    \State Generate matrix and broadcast data;\label{algline:BBFGBroadcast}
    \Else
    \State Receive data from the process 0;\label{algline:BBFGReceive}
    \EndIf
    \State $length=\lceil (2^{n-1}-1)/P\rceil$;
    \State $start=rank*length$;
    \State $end=\min (start+length,2^{n-1}-1)$;
    \For{$iter=start$ to $end$}\Comment{Parallelized using OpenMP}\label{algline:BBFGLoopStart}
    \State Generate set $\delta=\{\delta_i\}$ ($\delta_i=1$ if the $i^\text{th}$ bit of $iter$ is 1, $\delta_i=-1$ otherwise);\label{algline:BBFGLoopGenSet}
    \State $Per=Per+(\prod_{k=1}^n\delta_k)\prod_{j=1}^n\sum_{i=1}^n\delta_i a_{ij}$;\label{algline:BBFGLoopCalPer}
    \EndFor\label{algline:BBFGLoopEnd}
    \State Sum $Per$s from all processes \& store result on process 0;\label{algline:BBFGReduce}
    \If{$rank==0$}
    \State $Per=Per/2^{n-1}$;
    \EndIf
    \State Process 0 outputs $Per$;
  \end{algorithmic}
\end{algorithm}

\subsection{Load Balance}

We optimized the parallelized Ryser's algorithm through load balance. In Algorithm~\ref{alg:ParallelRyserProg}, within the calculation of $\sum_{j \in S}a_{ij}$, the number of add operations is related to $|S|$. However, different $S$s generated from $iter$ in Algorithm~\ref{alg:ParallelRyserProg} may have different sizes, leading to an imbalance loads. Therefore, we designed an optimizing algorithm with load balance, as Algorithm~\ref{alg:LoadBalance} shows. However, BB/FG's algorithm is free from the imbalance problem because each iteration in the loop of Algorithm~\ref{alg:ParallelBBFGProg} always performs $n$ add/subtract operations.

\begin{algorithm}[htbp]
  \caption{\small Ryser's algorithm with load balance}
  \label{alg:LoadBalance}
  \begin{algorithmic}[1]
    \Require $A$: an $n\times n$ matrix; $P$: number of processes
    \Ensure $Per$: Permanent of A
    \State Initialization;\Comment{For parallelization using MPI}
    \State Set $rank$ to current process ID;
    \If{$rank==0$}
    \State Generate matrix and broadcast data;
    \Else
    \State Receive data from the process 0;
    \EndIf
    \State $length=\lceil (2^{n-1}-1)/P\rceil$;
    \State $start=rank*length$;
    \State $end=\min (start+end,2^{n-1}-1)$;
    \For{$iter=start$ to $end$}\Comment{Parallelized using OpenMP}
    \State Generate set $S$ ($i\in S$ iff $i^\text{th}$ bit of $iter$ is 1);
    \State $Per=Per+(-1)^{|S|}\prod_{i=1}^n\sum_{j \in S}a_{ij}$;
    \State Generate the complementary set $S_C$ of $S$;
    \State $Per=Per+(-1)^{|S_C|}\prod_{i=1}^n\sum_{j \in S_C}a_{ij}$;
    \EndFor
    \State Sum $Per$s from all processes \& store result on process 0;
    \If{$n$ is odd \textbf{and} $rank==0$}
    \State $Per=-1*Per$;
    \EndIf
    \State Process 0 outputs $Per$;
  \end{algorithmic}
\end{algorithm}

\begin{algorithm}[htbp]
  \caption{\small Vectorized BB/FG's algorithm on MIC}
  \label{alg:BBFGMIC}
  \begin{algorithmic}[1]
    \Require $A$: an $n\times n$ matrix
    \Ensure $Per$: Permanent of A
    \State Set $vsize$ to the length of a vector register;
    \For{$iter=0$ to $2^{n-1}-1$}\label{algline:BBFGMICLoopStart}
    \State Generate set $\delta=\{\delta_i\}$ ($\delta_i=1$ if the $i^\text{th}$ bit of $iter$ is 1, $\delta_i=-1$ otherwise);\label{algline:BBFGMICGenSet}
    \State Set vector $(a_{ij})_V$ to $\left(a_{i,(j-1)\cdot vsize+1},\ldots a_{i,j\cdot vsize}\right)$;\label{algline:BBFGMICSetVec}
    \State $(vt)_V=\prod_{j=1}^{n/vsize}\sum_{i=1}^n\delta_i (a_{ij})_V$;\Comment{Vector operations}\label{algline:BBFGMICVecOp}
    \State Multiplicative reduction on vector $(vt)_V$ \& store result on  number $t$;\label{algline:BBFGMICMulRed}
    \State $Per=Per+(\prod_{k=1}^n\delta_k)\cdot t$;\label{algline:BBFGMICCalPer}
    \EndFor\label{algline:BBFGMICLoopEnd}
    \State $Per=Per/2^{n-1}$;
    \State Output $Per$;
  \end{algorithmic}
\end{algorithm}

\begin{algorithm}[htbp]
  \caption{\small Heterogeneous BB/FG's algorithm}
  \label{alg:BBFGheterogeneous}
  \begin{algorithmic}[1]
    \Require $A$: an $n\times n$ matrix; $P$: number of processes
    \Ensure $Per$: Permanent of A
    \State Initialization;\Comment{For parallelization using MPI}
    \State Dispatch computing tasks between CPUs and MICs;
    \State Sum $Per$s from MICs \& store result on host;
    \State Sum $Per$s from all processes \& store result on process 0;
    \State Process 0 computes $Per=Per/2^{n-1}$ \& outputs $Per$;
  \end{algorithmic}
\end{algorithm}

\subsection{Utilizing both CPUs and MICs}

To make full use of computing power of Tianhe-2, we implemented a vectorized BB/FG's algorithm for utilizing MIC accelerators. After several optimization steps, shown in TABLE~\ref{TAB:OPTVersion}, the final BB/FG's algorithm is Algorithm~\ref{alg:BBFGMIC}. Finally, we combined all optimization techniques and implemented a heterogeneous algorithm, Algorithm~\ref{alg:BBFGheterogeneous}, for utilizing all CPUs and accelerators of Tianhe-2.

\begin{table}[htbp]
  \caption{\footnotesize{\textbf{History of optimizations for co-processors.} Some attempts that are not efficient are not listed, such as using mask operation for completing $\delta$ related computation.}}\label{TAB:OPTVersion}
  \centering
  \begin{tabular}{clc}
  \hline
  Program Version & \multicolumn{1}{c}{Optimization steps} & Speedup \\
  \hline
  1 & Naive implementation & $1\times$ \\
  2 & Vectorization of matrix operation ($\sum_{i=1}^N\delta_ia_{ij}$) & $\sim$$1.6\times$\\
  3 & Vectorization of multiplicative reduction & $\sim$$1.8\times$\\
  4 & Multi-thread optimization & $\sim$$100\times$\\
  5 & Multi-MIC optimization & $\sim$$300\times$\\
  6 & Heterogeneous optimization with CPUs \& MICs & $\sim$$370\times$\\
  7 & Optimized the vectorization scheme & $\sim$$500\times$\\
  \hline
  \end{tabular}
\end{table}

\subsection{Efficiency}

\begin{figure}[htbp]
  \centering
  \includegraphics[width=.9\linewidth]{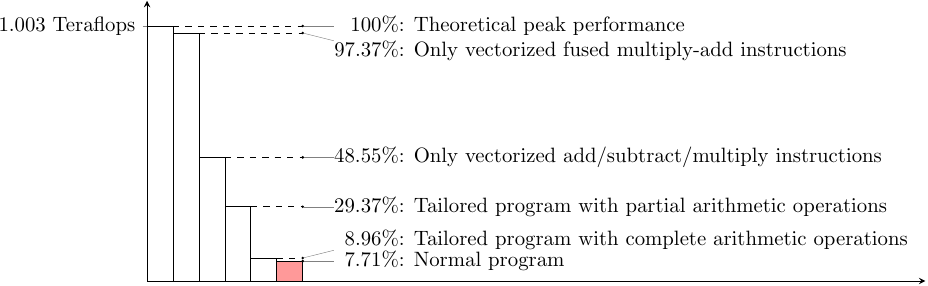}
  \caption{\footnotesize\textbf{Efficiency for utilizing co-processors.}
  The theoretical peak performance of a co-processor, Intel Xeon Phi, is 1.003 Teraflops. The performance of programs with only vectorized arithmetic instructions shows the capacity bound of the co-processor in real tests regardless of applications. The performance of two tailored programs takes the features of applications into consideration.}
  \label{fig:efficiencyMIC}
\end{figure}

The optimized program for co-processors has a performance of 77.4 Gigaflops, about 7.71\% of the theoretical peak performance. To evaluate the efficiency in more details, we tested several programs to put the co-processors into real tests.

As shown in FIG.~\ref{fig:efficiencyMIC}, a program with only vectorized fused multiply-add instructions has a performance of 97.37\% of the theoretical peak performance, while that with only add/subtract/multiply instructions has an efficiency of 48.55\%.

The tests with only vectorized arithmetic instructions do not take the features of applications into consideration. To evaluate how efficiently we have been utilizing CPUs and co-processors for BB/FG's algorithm, we tailored the program by removing implementation-related instructions as many as possible. The tailored programs, in TABLE~\ref{TAB:ProgramVersions}, are viewed as the baselines. We compared several implementations for $\delta$-related calculation in Equation~(\ref{eq:bbfg}) and adopted that, using branch instructions, with the highest efficiency in the final program. The program labelled ``complete'' in TABLE~\ref{TAB:ProgramVersions} contains these branch instructions, while the program labelled ``partial'' doesn't. If we take the tailored program with complete arithmetic operations left as a baseline, the average efficiencies are 97.0\% and 86.0\% for CPUs and co-processors respectively. If we take the program with partial arithmetic operations left, the average efficiencies are 74.0\% and 26.3\%.

\begin{table}[htbp]
  \caption{\footnotesize\textbf{Programs for efficiency evaluation.} ``Normal'' denotes the normal program. ``Complete'' denotes the tailored program labelled ``program consisting of complete arithmetic operations'' in FIG.~\ref{fig:efficiency}, and ``partial'' denotes that labelled ``program consisting of partial arithmetic operations'' in FIG.~\ref{fig:efficiency}. The programs for co-processors are vectorized. Only loop parts of the programs, line~\ref{algline:BBFGLoopStart}-\ref{algline:BBFGLoopEnd} in Algorithm~\ref{alg:ParallelBBFGProg} and line~\ref{algline:BBFGMICLoopStart}-\ref{algline:BBFGMICLoopEnd} in Algorithm~\ref{alg:BBFGMIC}, are discussed here.}\label{TAB:ProgramVersions}
  \centering\setlength\tabcolsep{5pt}
  \begin{tabular}{lll}
  \hline
  Program & \multicolumn{1}{c}{Instructions contained} & \multicolumn{1}{c}{Related equations} \\
  \hline
  \multirow{4}{*}{Normal}
   & (vectorized) add/subtract instructions & \multirow{4}{*}{$\sum_\delta\left(\prod_{k=1}^n\delta_k\right)\prod_{i=1}^n\sum_{j=1}^n\delta_ja_{ij}$} \\
   & (vectorized) multiply instructions & \\
   & branch instructions & \\
   & (vectorized) load/store instructions & \\
  \hline
  \multirow{3}{*}{Complete}
   & (vectorized) add/subtract instructions & \multirow{3}{*}{$\sum_\delta\left(\prod_{k=1}^n\delta_k\right)\prod_{i=1}^n\sum_{j=1}^n\delta_ja_{ij}$} \\
   & (vectorized) multiply instructions & \\
   & branch instructions & \\
  \hline
  \multirow{2}{*}{Partial}
   & (vectorized) add/subtract instructions & \multirow{2}{*}{$\sum_\delta\prod_{i=1}^n\sum_{j=1}^na_{ij}$} \\
   & (vectorized) multiply instructions & \\
  \hline
  \end{tabular}
\end{table}

\subsection{Fitting}

The fitting results can be found in TABLE~\ref{tab:RyserFittingResult} and TABLE~\ref{tab:BBFGFittingResult} for Ryser's algorithm and BB/FG's algorithm respectively.

\begin{table*}[htbp]
  \centering
  \caption{\footnotesize{\textbf{Fitting Result for Ryser's algorithm.}
      This table shows the fitting result for Ryser's algorithm in FIG.~\ref{fig:executionTime}(C).}}
  \label{tab:RyserFittingResult}
  \begin{tabular}{rccc}
    \hline
    \multicolumn{4}{l}{\bf Result:} \\
    \hline
    \multirow{2}{*}{\bf Coefficients} & \multirow{2}{*}{\bf Value} & \multicolumn{2}{c}{\bf 95\% confidence bounds} \\
    \cline{3-4}
    & & {\bf lower bound} & {\bf upper bound} \\
    \hline
    {\bf a:} & $5.319\times 10^{-11}$ & $4.135\times 10^{-11}$ & $6.502\times 10^{-11}$ \\
    {\bf b:} & $0.9924$ & $0.9670$ & $1.0170$ \\
    \hline
    \multicolumn{4}{l} {\bf Goodness of fitting:} \\
    \hline
    {\bf SSE:} & 29120 & \multicolumn{2}{c}{The smaller, the better} \\
    {\bf R\_square:} & 0.9996 & \multicolumn{2}{c}{\multirow{2}{*}{The closer to 1, the better} } \\
    {\bf Adjusted R\_square:} & 0.9996 & \multicolumn{2}{c}{}\\
    {\bf RMSE:} & 39.15 &\multicolumn{2}{c}{The smaller, the better} \\
    \hline
  \end{tabular}
\end{table*}

\begin{table*}[htbp]
  \centering
  \caption{\footnotesize{\textbf{Fitting Result for BB/FG's algorithm.}
      This table shows the fitting result for BB/FG's algorithm in FIG.~\ref{fig:executionTime}(D).}}
  \label{tab:BBFGFittingResult}
  \begin{tabular}{rccc}
    \hline
    \multicolumn{4}{l}{\bf Result:} \\
    \hline
    \multirow{2}{*}{\bf Coefficients} & \multirow{2}{*}{\bf Value} & \multicolumn{2}{c}{\bf 95\% confidence bounds} \\
    \cline{3-4}
    & & {\bf lower bound} & {\bf upper bound} \\
    \hline
    {\bf a:} & $9.805\times 10^{-12}$ & $9.245\times 10^{-12}$ & $1.037\times 10^{-11}$  \\
    {\bf b:} & $0.87782$ & $0.8916$ & $0.8649$ \\
    \hline
    \multicolumn{4}{l} {\bf Goodness of fitting:} \\
    \hline
    {\bf SSE:} & 284.2 & \multicolumn{2}{c}{The smaller, the better} \\
    {\bf R\_square:} & 0.994 & \multicolumn{2}{c}{\multirow{2}{*}{The closer to 1, the better} } \\
    {\bf Adjusted R\_square:} & 0.9938 & \multicolumn{2}{c}{}\\
    {\bf RMSE:} & 2.665 &\multicolumn{2}{c}{The smaller, the better} \\
    \hline
  \end{tabular}
\end{table*}

\subsection{ Architecture Transition}

Co-processors provide strong processing power for state-of-the-art supercomputers. The co-processors, Intel Xeon Phi 31S1P, in Tianhe-2 were released in 2013. In this section, we compare it with some new co-processors as shown in TABLE~\ref{tab:architectureTransition}.

For our program, there are three main aspects that may affect the performance obtained from co-processors. The first one is memory access performance due to the memory wall problem. Latency and bandwidth are two metrics associated with it. As shown in FIG.~\ref{fig:efficiencyMIC}, the tailored program with complete arithmetic operations, from which memory access instructions have been removed, has a small speedup, $\sim$1.16$\times$. Meanwhile, the profiling shows that a memory bandwidth of only $\sim$0.3 GB/s is requested by our program. These indicate that our program is compute intensive and its performance is mainly dominated by the computing part. The second aspect is multi-thread optimization for a single core. In Intel Xeon Phi 31S1P, each core has four threads. The tests show that multi-thread optimization can utilize the vector processing unit in each core with as high efficiency as possible. The last one is multi-thread optimization for multiple cores. Our program has only one inter-thread communication in the final phase of each thread, which is a reduction operation shown in line~\ref{algline:BBFGMICCalPer} of Algorithm~\ref{alg:BBFGMIC}. This brings the good scalability of multi-thread optimizations.

The Knights Landing microarchitecture integrates on-package memory for significantly higher memory bandwidth. For example, Intel Xeon Phi 7290 has a MCDRAM (Multi-Channel DRAM) bandwidth of $400+$ GB/s. Besides, the peak double precision performance of single core and whole co-processor are both upgraded. Even though memory bandwidth is not the bottleneck of our program, we still could expect optimistically a speedup of $\sim$$3\times$ on Intel Xeon Phi 7290 because of the advance of the peak performance.

NVIDIA Volta GV100 has much more performance than other co-processors. However, it adopts Volta, an NVIDIA-developed GPU microarchitecture. Compared to Intel Xeon Phi 31S1P, it has different microarchitecture and different programming model. Thus, it is hard to compare the performance of our program on NVIDIA Volta GV100 with that on Intel Xeon Phi 31S1P. Considering the feature of the algorithm, we could expect a good performance optimistically. Matrix-2000 is a 128-core co-processor. Each core has two 256-bit vector processing units. Just like the analysis for NVIDIA Volta GV100, we believe it is not hard to exploit both thread-parallelism and SIMD-parallelism of Matrix-2000. In conclusion, our program is compute intensive along with good scalability, and we could expect the transplant of our program to other architectures has a good performance.

\begin{table}[htbp]
  \caption{\footnotesize\textbf{Parameters of several latest co-processors.}
  Intel Xeon Phi 31S1P released in 2013 is the co-processor of Tianhe-2. Intel Xeon Phi 7290, using MIC architecture, is released in 2016. NVIDIA Volta GV100 released in 2018 is used in several new supercomputers. Matrix-2000 released in 2017 is used in Tianhe-2A, an upgraded system.}
  \label{tab:architectureTransition}
  \begin{tabular}{c|cccc}
    \hline
    Co-processor & Intel Xeon Phi 31S1P & Intel Xeon Phi 7290 & NVIDIA Volta GV100 & NUDT Matrix-2000 \\
    \hline
    Microarchitecture & Knights Corner & Knights Landing & Volta & Matrix-2000 \\
    Release year & 2013 & 2016 & 2018 & 2017 \\
    \# of cores & 57 & 72 & CUDA: 5120 \& Tensor: 640 & 128 \\
    \# of threads & 228 & 288 & - & 128 \\
    Clock (GHz) & 1.1 & 1.5 (Turbo: 1.7) & 1.132 (Boost: 1.628) & 1.2 \\
    Memory bandwidth (GB/s) & 320 & 102.4 & 870 & 143.1 \\
    MCDRAM bandwidth (GB/s) & - & 400+ & - & - \\
    Peak DP compute (Teraflops) & 1.003 & 3.456 & 7.400 & 2.458 \\
    \hline
  \end{tabular}
\end{table}

\section{Analysis of the Precision Issue}

The precision issue comes from the accumulated rounding errors introduced by limited word length of classical computers. Intermediate and final results are stored in double-precision floating-point format of IEEE-754 standard, where the total precision is decided by the 52-bit significand and an implicit bit. These 53 bits are approximately 16 decimal digits. Ryser's algorithm may produce intermediate result that is extremely larger than the final result, so that the most important bits were used for the intermediate result, and the final result becomes as large as what could be truncated. For example, in the case when computing the permanent of a $30\times 30$ all-one matrix, the largest intermediate result is $30^{30}\approx 2.06\times10^{44}$ while the final result is $30!\approx 2.65\times 10^{32}$ that is $10^{-12}$ smaller than the intermediate result. In these cases the errors of Ryser's algorithm may not be omitted. For BB/FG's algorithm, the final division guarantees the final result is in the same order of the intermediate data. Thus the error of BB/FG's algorithm accumulates much slower, so that the results from BB/FG's algorithm is much trustable than that produced by Ryser's.

To evaluate errors of the two algorithms, we computed absolute errors and relative error rates of permanents of all-one matrices with different $n$. As shown in FIG.~\ref{fig:errorsAllOneMatrices} and the detailed test data in TABLE~\ref{tab:ERRORBBFG} $\sim$ TABLE~\ref{tab:ERRORRyserComplex}, both the errors of Ryser's algorithm and BB/FG's algorithm grow exponentially with the increase of $n$ according to the fitting lines. However, the relative error rates of Ryser's algorithm reaches nealy 100\% when $n\geq 30$, and the errors of Ryser's algorithm are approximately $10^7 \sim 10^9$ larger than that of BB/FG's algorithm, while the the fit-line predicts the relative error rate of BB/FG's algorithm for a $50\times 50$ all-one matrix was about $0.0006\%$, and error rates does not exceed $10\%$ when the order of matrix is below 60. These results indicate that for relatively large $n$, the precision issue overburdens Ryser's algorithm and BB/FG's algorithm could still maintain the accuracy. Our results recommend BB/FG's algorithm for the classical rival of quantum Boson sampling in the future research when the experiment scales up to a certain size, rather than Ryser's algorithm that most considered before.

Realistically, the randomness of the built matrices make their permanents hard to evaluate. To confirm the precision issue in more realistic situations, we generated three types of random matrices, as shown in FIG.~\ref{fig:errorsRandomMatrix}, and compare the results of the two algorithms. The errors of random unitary matrices and randomly derived matrices were marginal, thus we believe these results are trustable. But that of specially derived matrix chosen were still very large. The growing speed of errors of specially derived matrices is exponential. This suggests a double check with both two algorithms may be necessary for future research to verify results produced by classical computers. Besides, the precision issue implicates quantum computation may outperforms classical computation not just in speed, but also in precision sometimes.

\section{Detailed Test Data}
\label{app:detailedTestData}

\subsection{Speed Performance}
\label{app:speedPerformance}

TABLE~\ref{tab:preTestRYSER} and TABLE~\ref{tab:preTestBBFG} show the detailed test data of Ryser's algorithm and BB/FG's algorithm respectively. Some of data in these tables is used in FIG.~\ref{fig:executionTime} (main text).

\begin{figure*}
  \centering
  \includegraphics[angle=270,width=\linewidth]{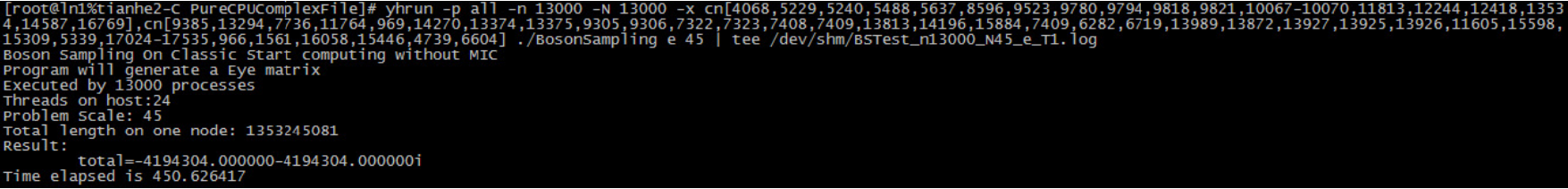}
  \caption{\footnotesize{\textbf{Screenshot of computing the  permanent of a $45 \times 45$ matrix on 13,000 nodes.}
      The command line is ``yhrun -p all -n 13000 -N 13000 -x cn[4068...6604] ./BosonSampling e 45 $|$ tee /dev/shm/BSTest\_n13000\_N45\_e\_T1.log'', where ``yhrun'' is the command to submit the task, ``-p all'' means the program would be executed on all the partitions of Tianhe-2, ``-n 13000 -N 13000'' means we would start 13,000 processes on 13,000 computing nodes, ``-x cn[4068...6604]'' defines the nodes that we won't use for this computation and ``./BosonSampling e 45'' shows the input is a $45 \times 45$-matrix with diagonal elements to be $1+i$ and other elements to be 0. The output would be recorded in file ``/dev/shm/BSTest\_n13000\_N45\_e\_T1.log''. The line after ``Result:'' gives the computing result, which equals to the theoretical value $(1+i)^{45}$, and the last line gives the execution time in unit of seconds.
    }}
  \label{fig:ScreenshotMatrix45Nodes13000}
\end{figure*}

\begin{figure*}
  \centering
  \includegraphics[angle=270,width=\linewidth]{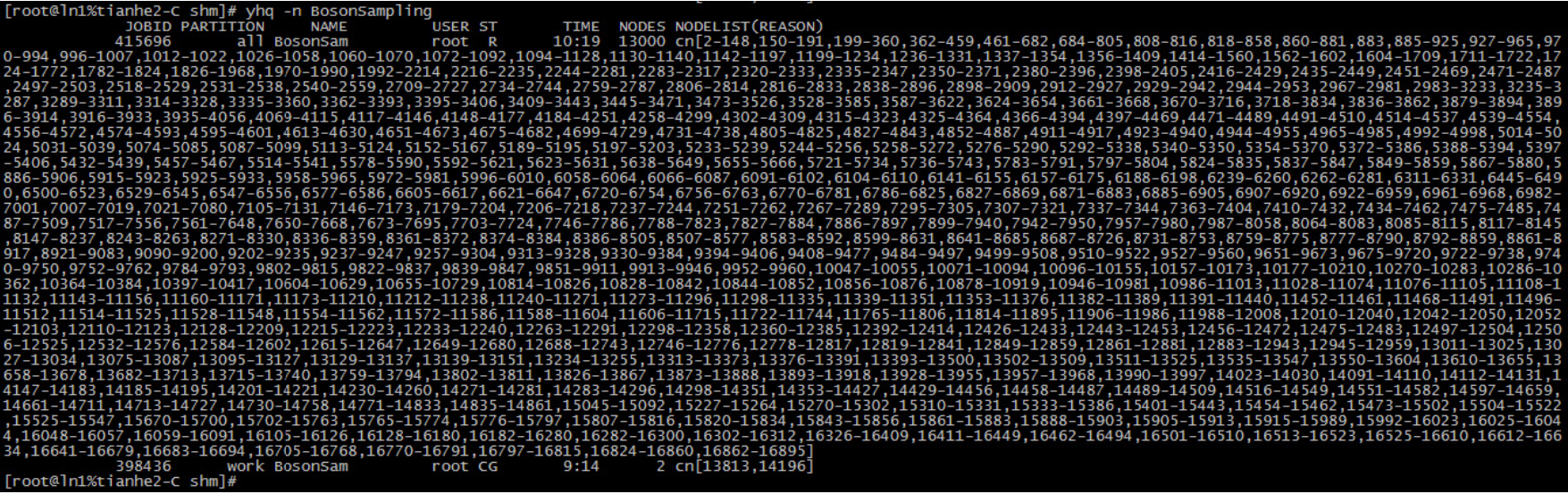}
  \caption{\footnotesize{\textbf{Screenshot of the node list of the 13,000 nodes used.}
      The command line is ``yhq -n BosonSampling'', which output the job with the executive named ``BosonSampling'' in the job sequence, the partition where this job was allocated and above all, and the node list of this job.
    }}
  \label{fig:Screenshot13000NodeList}
\end{figure*}

\begin{figure*}
  \centering
  \includegraphics[angle=270,width=.45\linewidth]{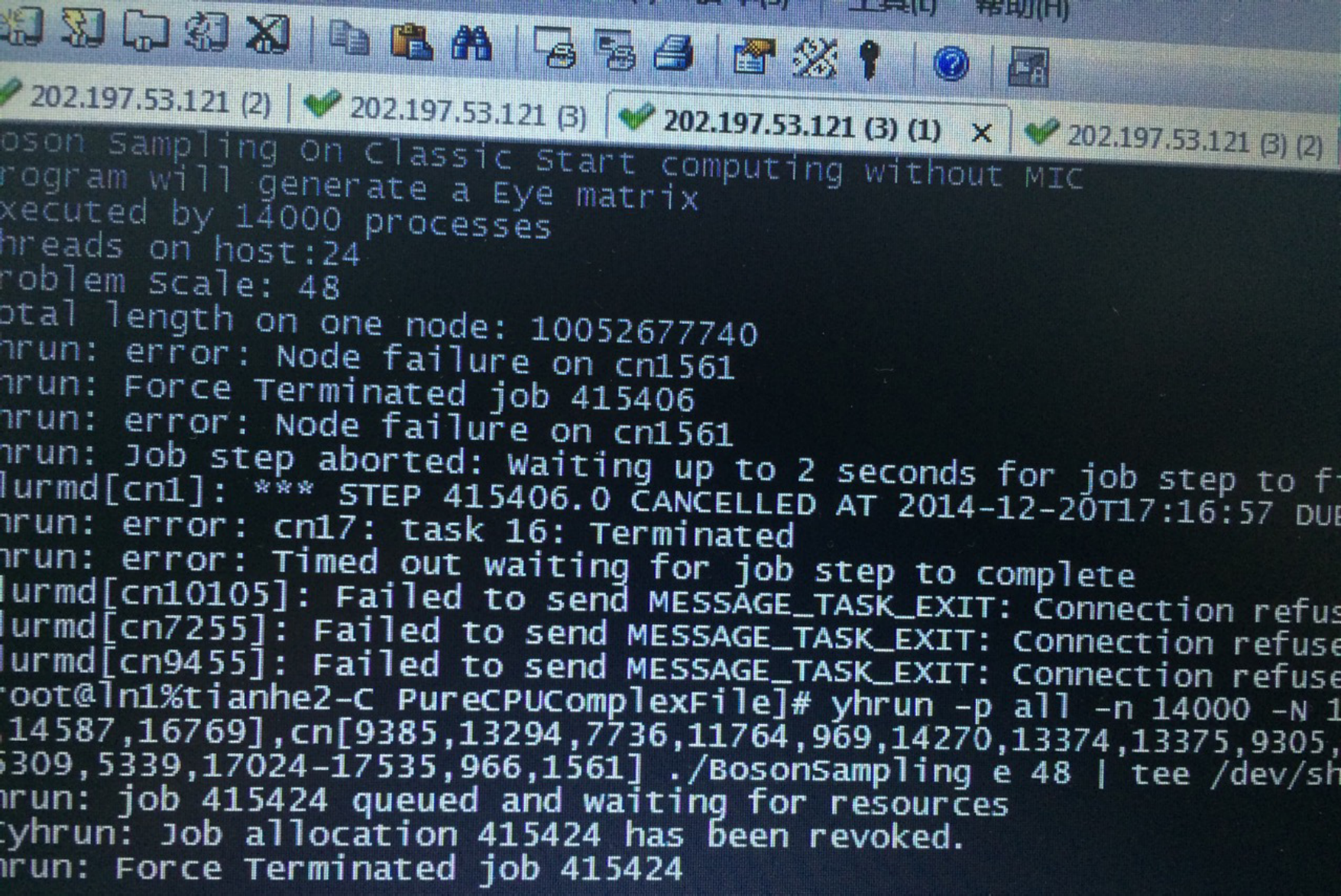}
  \caption{\footnotesize{\textbf{Photograph that shows error occurred during the computation on 14,000 nodes.}
      Errors occurred more frequently when the number of nodes increases~\protect\cite{Yang2012s}. In this photograph, the node cn1561 failed during the execution, leading to the job termination and more errors in the node communication.
    }}
  \label{fig:PhotographNodes14000Error}
\end{figure*}

Some screenshots and a photograph are shown here. FIG.~\ref{fig:ScreenshotMatrix45Nodes13000} and FIG.~\ref{fig:Screenshot13000NodeList} are screenshots in which Tianhe-2 was computing the permanent of a $45\times 45$ matrix. FIG.~\ref{fig:PhotographNodes14000Error} shows an case of error in which some node failed during the execution.

\subsection{Precision Performance}
\label{app:precisionPerformance}

TABLE~\ref{tab:ERRORBBFG} $\sim$ TABLE~\ref{tab:ERRORRyserComplex} show the detailed test errors of BB/FG's algorithm and Ryser's algorithm. TABLE~\ref{tab:ERRORBBFG} and TABLE~\ref{tab:ERRORRyser} show results for real matrices, and some of data is used in FIG.~\ref{fig:errorsAllOneMatrices} (main text). TABLE~\ref{tab:ERRORBBFGComplex} and TABLE~\ref{tab:ERRORRyserComplex} show results for complex matrices.

Performing add operations in different orders may produce different intermediate results, which may bring different accumulated rounding errors. We tested Ryser's algorithm with different orders of add operations. As shown in TABLE~\ref{tab:ErrorOrder}, we found the precision issue of Ryser's algorithm became worse in some cases.

TABLE~\ref{tab:ResultsComparing}, TABLE~\ref{tab:ResultsComparingDerived} and TABLE~\ref{tab:SpecialCase} show the detailed test errors for random matrices, and some of data is used in FIG.~\ref{fig:errorsRandomMatrix} (main text).

\bibliographystyle{unsrt}

\begin{table*}
  \centering
  \caption{\bf Performance Parameters of Tianhe-2 Supercomputer.}
  \label{tab:PerfPara}
  \begin{tabular}{cccc}
    \hline
    \multicolumn{2}{c}{{\bf Item}} & {\bf Parameters of a node} &{\bf Parameters of the System} \\
    \hline
    \multicolumn{2}{c}{Peak Performance} & 3.43 Teraflops & 54.90 Petaflops \\
    \hline
    \multirow{2}{*}{Processor} & Intel Xeon E5 (12 cores) & 2 (24 cores) & 32,000 (384 thousand cores) \\ \cline{2-4}
    & Xeon Phi (57 cores) & 3 (171cores) & 48,000 (2.736 million Cores) \\
    \hline
    \multicolumn{2}{c}{Memory Storage Capacity} & 64GB+8GB (Xeon Phi) & 1.408PB\\
    \hline
    \multicolumn{2}{c}{Disk Capacity} & \multicolumn{2}{c}{12.4PB}\\
    \hline
    \multicolumn{2}{c}{Mainboard (Two computing nodes)} & \multicolumn{2}{c}{8,000}\\
    \hline
    \multicolumn{2}{c}{Front-Ending Processor FT-1500(16 cores)} & \multicolumn{2}{c}{4,096}\\
    \hline
    \multicolumn{2}{c}{Interconnect Network} & \multicolumn{2}{c}{TH Express-2}\\
    \hline
    \multicolumn{2}{c}{Operating System} & \multicolumn{2}{c}{Kylin}\\
    \hline
    \multicolumn{2}{c}{Power Consumption} & \multicolumn{2}{c}{24 MW (17.808 MW without cooling system)}\\
    \hline
  \end{tabular}
\end{table*}

\begin{table*}
  \caption{\footnotesize{\textbf{Execution time of Ryser's algorithm (with only CPUs).}
      The data of ``Ryser@CPU'' in FIG.~\ref{fig:executionTime} comes from this table. We used $P$ computing nodes to compute permanents of $n\times n$ matrices.}}
  \centering
  \label{tab:preTestRYSER}
  \begin{tabular}{ccr|ccr|ccr}
    \hline
    \multicolumn{1}{p{1cm}}{\centering $P$} & \multicolumn{1}{p{1cm}}{\centering $n$} & Execution time (s) & \multicolumn{1}{p{1cm}}{\centering $P$} & \multicolumn{1}{p{1cm}}{\centering $n$} & Execution time (s) & \multicolumn{1}{p{1cm}}{\centering $P$} & \multicolumn{1}{p{1cm}}{\centering $n$} & Execution time (s) \\
    \hline
    1  &   25  &   1.9642$\pm$0.2372\hspace{5pt}  &   2  &   26  &   2.1562$\pm$0.2068\hspace{5pt}  &   4  &   27  &   2.3908$\pm$0.1985\hspace{5pt}  \\
    1  &   26  &   3.7839$\pm$0.2816\hspace{5pt}  &   2  &   27  &   4.293$\pm$0.4017\hspace{5pt}  &   4  &   28  &   4.7673$\pm$0.3210\hspace{5pt}  \\
    1  &   27  &   7.6726$\pm$0.5345\hspace{5pt}  &   2  &   28  &   8.6394$\pm$0.4483\hspace{5pt}  &   4  &   29  &   9.3054$\pm$0.5000\hspace{5pt}  \\
    1  &   28  &   15.9046$\pm$1.1006\hspace{5pt}  &   2  &   29  &   17.6308$\pm$1.5244\hspace{5pt}  &   4  &   30  &   18.8293$\pm$1.1771\hspace{5pt}  \\
    1  &   29  &   31.8353$\pm$1.5959\hspace{5pt}  &   2  &   30  &   34.267$\pm$2.0059\hspace{5pt}  &   4  &   31  &   37.5951$\pm$2.0244\hspace{5pt}  \\
    1  &   30  &   68.1029$\pm$3.5834\hspace{5pt}  &   2  &   31  &   71.9743$\pm$1.2699\hspace{5pt}  &   4  &   32  &   75.6077$\pm$0.8866\hspace{5pt}  \\
    1  &   31  &   143.1606$\pm$3.0337\hspace{5pt}  &   2  &   32  &   148.82$\pm$0.4387\hspace{5pt}  &   4  &   33  &   159.1373$\pm$3.7312\hspace{5pt}  \\
    8  &   28  &   2.7479$\pm$0.1168\hspace{5pt}  &   16  &   29  &   2.8932$\pm$0.1206\hspace{5pt}  &   32  &   30  &   3.2313$\pm$0.0400\hspace{5pt}  \\
    8  &   29  &   5.3319$\pm$0.4006\hspace{5pt}  &   16  &   30  &   5.7524$\pm$0.2777\hspace{5pt}  &   32  &   31  &   6.4901$\pm$0.1221\hspace{5pt}  \\
    8  &   30  &   9.803$\pm$0.3052\hspace{5pt}  &   16  &   31  &   11.7995$\pm$1.0153\hspace{5pt}  &   32  &   32  &   12.5818$\pm$1.0329\hspace{5pt}  \\
    8  &   31  &   20.4626$\pm$0.8248\hspace{5pt}  &   16  &   32  &   22.9742$\pm$1.7605\hspace{5pt}  &   32  &   33  &   23.8788$\pm$1.6136\hspace{5pt}  \\
    8  &   32  &   39.3735$\pm$1.5259\hspace{5pt}  &   16  &   33  &   43.5551$\pm$1.7533\hspace{5pt}  &   32  &   34  &   47.7301$\pm$1.3930\hspace{5pt}  \\
    8  &   33  &   79.6846$\pm$1.0417\hspace{5pt}  &   16  &   34  &   87.3021$\pm$5.3410\hspace{5pt}  &   32  &   35  &   93.319$\pm$6.8016\hspace{5pt}  \\
    8  &   34  &   165.1047$\pm$1.9254\hspace{5pt}  &   16  &   35  &   176.1717$\pm$2.0559\hspace{5pt}  &   32  &   36  &   186.7612$\pm$5.0536\hspace{5pt}  \\
    64  &   31  &   3.4116$\pm$0.0873\hspace{5pt}  &   128  &   32  &   3.5867$\pm$0.0201\hspace{5pt}  &   256  &   33  &   3.9325$\pm$0.2320\hspace{5pt}  \\
    64  &   32  &   6.708$\pm$0.2296\hspace{5pt}  &   128  &   33  &   7.0565$\pm$0.1693\hspace{5pt}  &   256  &   34  &   7.8038$\pm$0.5269\hspace{5pt}  \\
    64  &   33  &   13.8701$\pm$1.8249\hspace{5pt}  &   128  &   34  &   16.0373$\pm$2.2439\hspace{5pt}  &   256  &   35  &   16.0211$\pm$1.0699\hspace{5pt}  \\
    64  &   34  &   25.9721$\pm$1.7686\hspace{5pt}  &   128  &   35  &   28.4909$\pm$1.8840\hspace{5pt}  &   256  &   36  &   30.9905$\pm$1.4728\hspace{5pt}  \\
    64  &   35  &   51.6904$\pm$5.1145\hspace{5pt}  &   128  &   36  &   54.5887$\pm$6.6629\hspace{5pt}  &   256  &   37  &   59.1407$\pm$4.2963\hspace{5pt}  \\
    64  &   36  &   98.3382$\pm$1.9439\hspace{5pt}  &   128  &   37  &   103.0265$\pm$5.2002\hspace{5pt}  &   256  &   38  &   108.5136$\pm$3.4498\hspace{5pt}  \\
    64  &   37  &   190.9616$\pm$4.0806\hspace{5pt}  &   128  &   38  &   203.6487$\pm$9.3798\hspace{5pt}  &   256  &   39  &   210.7175$\pm$12.3217\hspace{5pt}  \\
    \hline
  \end{tabular}
\end{table*}

\begin{table*}
  \caption{\footnotesize{\textbf{Execution time of heterogeneous BB/FG's algorithm (with both CPUs and MICs).}
      The data of ``BB/FG@Hybrid'' in FIG.~\ref{fig:executionTime} and the fitting data in FIG.~\ref{fig:executionTime}(C) come from this table. We used $P$ computing nodes to compute permanents of $n\times n$ matrices.
    }}
  \centering
  \label{tab:preTestBBFG}
  \begin{tabular}{ccc|ccc|ccc}
    \hline
    \multicolumn{1}{p{1cm}}{\centering $P$} & \multicolumn{1}{p{1cm}}{\centering $n$} & Execution time (s) & \multicolumn{1}{p{1cm}}{\centering $P$} & \multicolumn{1}{p{1cm}}{\centering $n$} & Execution time (s) & \multicolumn{1}{p{1cm}}{\centering $P$} & \multicolumn{1}{p{1cm}}{\centering $n$} & Execution time (s) \\
    \hline
            1  &   24  &   1.2384  &   2  &   25  &   1.0713  &   4  &   26  &   1.4582  \\
            1  &   25  &   1.0907  &   2  &   26  &   1.1171  &   4  &   27  &   1.5171  \\
            1  &   26  &   1.3009  &   2  &   27  &   1.4052  &   4  &   28  &   1.8441  \\
            1  &   27  &   1.6853  &   2  &   28  &   1.8302  &   4  &   29  &   2.3602  \\
            1  &   28  &   2.5140  &   2  &   29  &   2.8678  &   4  &   30  &   3.6961  \\
            1  &   29  &   4.3374  &   2  &   30  &   4.6822  &   4  &   31  &   6.1990  \\
            1  &   30  &   8.0175  &   2  &   31  &   9.0503  &   4  &   32  &   10.744  \\
            1  &   31  &   15.130  &   2  &   32  &   15.828  &   4  &   33  &   25.512  \\
            8  &   27  &   1.4967  &   16  &   28  &   1.5002  &   32  &   29  &   1.3799  \\
            8  &   28  &   1.6205  &   16  &   29  &   1.6593  &   32  &   30  &   1.6045  \\
            8  &   29  &   1.9648  &   16  &   30  &   1.9986  &   32  &   31  &   1.9624  \\
            8  &   30  &   2.6284  &   16  &   31  &   2.7252  &   32  &   32  &   2.7685  \\
            8  &   31  &   4.0077  &   16  &   32  &   4.1378  &   32  &   33  &   5.4228  \\
            8  &   32  &   6.4726  &   16  &   33  &   8.7461  &   32  &   34  &   9.9724  \\
            8  &   33  &   15.832  &   16  &   34  &   16.907  &   32  &   35  &   19.550  \\
            8  &   34  &   31.572  &   16  &   35  &   32.646  &   32  &   36  &   39.477  \\
            8  &   35  &   60.678  &   16  &   36  &   65.492  &   32  &   37  &   80.358  \\
    64 &         30 &    1.3939  &        128 &         31 &    1.4278  &        256 &         32 &    1.6232  \\

    64 &         31 &    1.5781  &        128 &         32 &    1.9457  &        256 &         33 &    1.9047  \\

    64 &         32 &    1.9385  &        128 &         33 &    2.1552  &        256 &         34 &    2.6021  \\

    64 &         33 &    3.3056  &        128 &         34 &    4.3292  &        256 &         35 &    4.5121  \\

    64 &         34 &    5.4466  &        128 &         35 &    5.9397  &        256 &         36 &    8.2317  \\

    64 &         35 &    11.228  &        128 &         36 &    15.533  &        256 &         37 &    15.877  \\

    64 &         36 &    21.428  &        128 &         37 &    23.652  &        256 &         38 &    31.815  \\

    64 &         37 &    44.767  &        128 &         38 &    46.065  &        256 &         39 &    66.710  \\

    64 &         38 &    84.098  &        128 &         39 &    133.18  &        256 &         40 &    132.35  \\
    \hline
  \end{tabular}
\end{table*}

\begin{table*}
  \caption{\footnotesize{\textbf{Errors of BB/FG's algorithm for real matrices.}
      Some data in FIG.~\ref{fig:errorsAllOneMatrices} comes from this table. We did the precision test by computing the permanent of an $n\times n$ all-$r$ real matrix, in which each element has a real value of $r$. In this case, the permanent in theory is $r^n\cdot n!$. In the table, $AbsErr=\left|Per_\text{BB/FG}-Per_\text{Theory}\right|$ and $RelErr=\left|(Per_\text{BB/FG}-Per_\text{Theory})/Per_\text{Theory}\right|$.
    }}
  \centering
  \label{tab:ERRORBBFG}
  \begin{tabular}{c|ccccc|ccccc}
    \hline
    $n$ & $r$ & $Per_\text{BB/FG}$ & $Per_\text{Theory}$ & $AbsErr$ & $RelErr$ & $r$ & $Per_\text{BB/FG}$ & $Per_\text{Theory}$ & $AbsErr$ & $RelErr$ \\
    \hline
    17 &          1 &   3.56E+14 &   3.56E+14 &   2.00E+00 &   5.62E-15 &        0.1 &   3.56E-03 &   3.56E-03 &   3.64E-17 &   1.02E-14 \\

    18 &          1 &   6.40E+15 &   6.40E+15 &   1.20E+02 &   1.87E-14 &        0.1 &   6.40E-03 &   6.40E-03 &   2.78E-16 &   4.35E-14 \\

    19 &          1 &   1.22E+17 &   1.22E+17 &   9.92E+02 &   8.15E-15 &        0.1 &   1.22E-02 &   1.22E-02 &   1.18E-15 &   9.67E-14 \\

    20 &          1 &   2.43E+18 &   2.43E+18 &   2.00E+04 &   8.21E-15 &        0.1 &   2.43E-02 &   2.43E-02 &   4.25E-15 &   1.75E-13 \\

    21 &          1 &   5.11E+19 &   5.11E+19 &   2.74E+06 &   5.36E-14 &        0.1 &   5.11E-02 &   5.11E-02 &   4.36E-15 &   8.54E-14 \\

    22 &          1 &   1.12E+21 &   1.12E+21 &   1.02E+08 &   9.11E-14 &        0.1 &   1.12E-01 &   1.12E-01 &   6.60E-14 &   5.87E-13 \\

    23 &          1 &   2.59E+22 &   2.59E+22 &   1.20E+10 &   4.63E-13 &        0.1 &   2.59E-01 &   2.59E-01 &   1.06E-13 &   4.08E-13 \\

    24 &          1 &   6.20E+23 &   6.20E+23 &   3.27E+11 &   5.28E-13 &        0.1 &   6.20E-01 &   6.20E-01 &   7.66E-13 &   1.23E-12 \\

    25 &          1 &   1.55E+25 &   1.55E+25 &   8.28E+12 &   5.34E-13 &        0.1 &   1.55E+00 &   1.55E+00 &   4.33E-12 &   2.79E-12 \\

    26 &          1 &   4.03E+26 &   4.03E+26 &   1.94E+15 &   4.81E-12 &        0.1 &   4.03E+00 &   4.03E+00 &   7.01E-12 &   1.74E-12 \\

    27 &          1 &   1.09E+28 &   1.09E+28 &   1.47E+17 &   1.35E-11 &        0.1 &   1.09E+01 &   1.09E+01 &   8.24E-11 &   7.57E-12 \\

    28 &          1 &   3.05E+29 &   3.05E+29 &   4.95E+17 &   1.62E-12 &        0.1 &   3.05E+01 &   3.05E+01 &   2.18E-10 &   7.15E-12 \\

    29 &          1 &   8.84E+30 &   8.84E+30 &   1.45E+20 &   1.64E-11 &        0.1 &   8.84E+01 &   8.84E+01 &   2.76E-09 &   3.12E-11 \\

    30 &          1 &   2.65E+32 &   2.65E+32 &   3.42E+22 &   1.29E-10 &        0.1 &   2.65E+02 &   2.65E+02 &   4.66E-09 &   1.76E-11 \\

    31 &          1 &   8.22E+33 &   8.22E+33 &   1.37E+24 &   1.67E-10 &        0.1 &   8.22E+02 &   8.22E+02 &   5.54E-08 &   6.73E-11 \\

    32 &          1 &   2.63E+35 &   2.63E+35 &   1.28E+26 &   4.87E-10 &        0.1 &   2.63E+03 &   2.63E+03 &   9.49E-07 &   3.61E-10 \\

    33 &          1 &   8.68E+36 &   8.68E+36 &   3.95E+27 &   4.55E-10 &        0.1 &   8.68E+03 &   8.68E+03 &   3.12E-07 &   3.60E-11 \\

    34 &          1 &   2.95E+38 &   2.95E+38 &   4.64E+28 &   1.57E-10 &        0.1 &   2.95E+04 &   2.95E+04 &   1.25E-05 &   4.25E-10 \\

    35 &          1 &   1.03E+40 &   1.03E+40 &   6.32E+30 &   6.12E-10 &        0.1 &   1.03E+05 &   1.03E+05 &   1.32E-04 &   1.28E-09 \\

    36 &          1 &   3.72E+41 &   3.72E+41 &   2.13E+33 &   5.72E-09 &        0.1 &   3.72E+05 &   3.72E+05 &   1.70E-03 &   4.56E-09 \\

    17 &         10 &   3.56E+31 &   3.56E+31 &   6.98E+17 &   1.96E-14 &        0.2 &   466.2066 &   466.2066 &   3.98E-12 &   8.53E-15 \\

    18 &         10 &   6.40E+33 &   6.40E+33 &   1.11E+20 &   1.73E-14 &        0.2 &   1678.344 &   1678.344 &   6.62E-11 &   3.94E-14 \\

    19 &         10 &   1.22E+36 &   1.22E+36 &   2.01E+22 &   1.65E-14 &        0.2 &   6377.707 &   6377.707 &   6.37E-10 &   9.98E-14 \\

    20 &         10 &   2.43E+38 &   2.43E+38 &   4.99E+24 &   2.05E-14 &        0.2 &   25510.83 &   25510.83 &   4.34E-09 &   1.70E-13 \\

    21 &         10 &   5.11E+40 &   5.11E+40 &   1.86E+27 &   3.63E-14 &        0.2 &   107145.5 &   107145.5 &   6.05E-09 &   5.65E-14 \\

    22 &         10 &   1.12E+43 &   1.12E+43 &   4.23E+29 &   3.77E-14 &        0.2 &   471440.1 &   471440.1 &   2.95E-07 &   6.25E-13 \\

    23 &         10 &   2.59E+45 &   2.59E+45 &   3.28E+32 &   1.27E-13 &        0.2 &    2168624 &    2168624 &   7.00E-07 &   3.23E-13 \\

    24 &         10 &   6.20E+47 &   6.20E+47 &   7.70E+35 &   1.24E-12 &        0.2 &   10409397 &   10409397 &   1.49E-05 &   1.43E-12 \\

    25 &         10 &   1.55E+50 &   1.55E+50 &   3.87E+38 &   2.50E-12 &        0.2 &   52046984 &   52046984 &   2.18E-04 &   4.19E-12 \\

    26 &         10 &   4.03E+52 &   4.03E+52 &   4.48E+40 &   1.11E-12 &        0.2 &   2.71E+08 &   2.71E+08 &   1.96E-04 &   7.22E-13 \\

    27 &         10 &   1.09E+55 &   1.09E+55 &   7.44E+43 &   6.84E-12 &        0.2 &   1.46E+09 &   1.46E+09 &   1.09E-02 &   7.44E-12 \\

    28 &         10 &   3.05E+57 &   3.05E+57 &   4.22E+46 &   1.38E-11 &        0.2 &   8.18E+09 &   8.18E+09 &   1.11E-01 &   1.36E-11 \\

    29 &         10 &   8.84E+59 &   8.84E+59 &   1.07E+50 &   1.21E-10 &        0.2 &   4.75E+10 &   4.75E+10 &   8.67E-01 &   1.83E-11 \\

    30 &         10 &   2.65E+62 &   2.65E+62 &   1.67E+52 &   6.31E-11 &        0.2 &   2.85E+11 &   2.85E+11 &   2.53E+01 &   8.87E-11 \\

    31 &         10 &   8.22E+64 &   8.22E+64 &   1.27E+55 &   1.55E-10 &        0.2 &   1.77E+12 &   1.77E+12 &   9.90E+01 &   5.61E-11 \\

    32 &         10 &   2.63E+67 &   2.63E+67 &   1.31E+58 &   4.97E-10 &        0.2 &   1.13E+13 &   1.13E+13 &   6.26E+03 &   5.54E-10 \\

    33 &         10 &   8.68E+69 &   8.68E+69 &   5.22E+60 &   6.01E-10 &        0.2 &   7.46E+13 &   7.46E+13 &   2.02E+04 &   2.71E-10 \\

    34 &         10 &   2.95E+72 &   2.95E+72 &   1.25E+62 &   4.23E-11 &        0.2 &   5.07E+14 &   5.07E+14 &   2.64E+05 &   5.20E-10 \\

    35 &         10 &   1.03E+75 &   1.03E+75 &   5.46E+66 &   5.28E-09 &        0.2 &   3.55E+15 &   3.55E+15 &   1.03E+07 &   2.91E-09 \\

    36 &         10 &   3.72E+77 &   3.72E+77 &   5.22E+68 &   1.40E-09 &        0.2 &   2.56E+16 &   2.56E+16 &   1.25E+08 &   4.87E-09 \\
    \hline
  \end{tabular}
\end{table*}

\begin{table*}
  \caption{\footnotesize{\textbf{Errors of BB/FG's algorithm for complex matrices.}
      We did the precision test by computing the permanent of an $n\times n$ all-$r$ complex matrix, in which each element has a complex value of $r$. In this case, the permanent in theory is $r^n\cdot n!$. $\text{Real}(Per_\text{BB/FG})$ is the real part of $Per_\text{BB/FG}$, and $\text{Imag}(Per_\text{BB/FG})$ is the imaginary part. $AbsErr=\left|Per_\text{BB/FG}-Per_\text{Theory}\right|$ and $RelErr=\left|(Per_\text{BB/FG}-Per_\text{Theory})/Per_\text{Theory}\right|$.}}
  \centering
  \label{tab:ERRORBBFGComplex}
  \begin{tabular}{cccccccc}
    \hline
    $n$ & $r$ & $\text{Real}(Per_\text{BB/FG})$ & $\text{Imag}(Per_\text{BB/FG})$ & $\text{Real}(Per_\text{Theory})$ & $\text{Imag}(Per_\text{Theory})$ & $AbsErr$ & $RelErr$ \\
    \hline
    17 &        $1+i$ &   9.11E+16 &   9.11E+16 &   9.11E+16 &   9.11E+16 &   4.30E+02 &   3.34E-15 \\

    18 &        $1+i$ &   0.00E+00 &   3.28E+18 &   0.00E+00 &   3.28E+18 &   7.01E+04 &   2.14E-14 \\

    19 &        $1+i$ &  -6.23E+19 &   6.23E+19 &  -6.23E+19 &   6.23E+19 &   1.14E+06 &   1.29E-14 \\

    20 &        $1+i$ &  -2.49E+21 &   0.00E+00 &  -2.49E+21 &   0.00E+00 &   1.99E+07 &   8.00E-15 \\

    21 &        $1+i$ &  -5.23E+22 &  -5.23E+22 &  -5.23E+22 &  -5.23E+22 &   4.10E+09 &   5.55E-14 \\

    22 &        $1+i$ &  -1.57E+10 &  -2.30E+24 &   0.00E+00 &  -2.30E+24 &   2.11E+11 &   9.14E-14 \\

    23 &        $1+i$ &   5.29E+25 &  -5.29E+25 &   5.29E+25 &  -5.29E+25 &   3.45E+13 &   4.61E-13 \\

    24 &        $1+i$ &   2.54E+27 &  -3.49E+12 &   2.54E+27 &   0.00E+00 &   1.34E+15 &   5.27E-13 \\

    25 &        $1+i$ &   6.35E+28 &   6.35E+28 &   6.35E+28 &   6.35E+28 &   4.78E+16 &   5.32E-13 \\

    26 &        $1+i$ &   1.49E+17 &   3.30E+30 &   0.00E+00 &   3.30E+30 &   1.60E+19 &   4.84E-12 \\

    27 &        $1+i$ &  -8.92E+31 &   8.92E+31 &  -8.92E+31 &   8.92E+31 &   1.70E+21 &   1.35E-11 \\

    28 &        $1+i$ &  -5.00E+33 &  -5.16E+19 &  -5.00E+33 &   0.00E+00 &   8.10E+21 &   1.62E-12 \\

    29 &        $1+i$ &  -1.45E+35 &  -1.45E+35 &  -1.45E+35 &  -1.45E+35 &   3.37E+24 &   1.64E-11 \\

    30 &        $1+i$ &  -7.79E+23 &  -8.69E+36 &   0.00E+00 &  -8.69E+36 &   1.12E+27 &   1.29E-10 \\

    31 &        $1+i$ &   2.69E+38 &  -2.69E+38 &   2.69E+38 &  -2.69E+38 &   6.36E+28 &   1.67E-10 \\

    32 &        $1+i$ &   1.72E+40 &   0.00E+00 &   1.72E+40 &   0.00E+00 &   8.40E+30 &   4.87E-10 \\

    33 &        $1+i$ &   5.69E+41 &   5.69E+41 &   5.69E+41 &   5.69E+41 &   3.66E+32 &   4.55E-10 \\

    34 &        $1+i$ &   0.00E+00 &   3.87E+43 &   0.00E+00 &   3.87E+43 &   6.09E+33 &   1.57E-10 \\

    35 &        $1+i$ &  -1.35E+45 &   1.35E+45 &  -1.35E+45 &   1.35E+45 &   1.36E+40 &   7.12E-06 \\

    36 &        $1+i$ &  -9.75E+46 &  -1.13E+35 &  -9.75E+46 &   0.00E+00 &   5.58E+38 &   5.72E-09 \\

    17 &     $10+10i$ &   9.11E+33 &   9.11E+33 &   9.11E+33 &   9.11E+33 &   2.40E+20 &   1.86E-14 \\

    18 &     $10+10i$ &   0.00E+00 &   3.28E+36 &   0.00E+00 &   3.28E+36 &   6.02E+22 &   1.84E-14 \\

    19 &     $10+10i$ &  -6.23E+38 &   6.23E+38 &  -6.23E+38 &   6.23E+38 &   2.13E+25 &   2.41E-14 \\

    20 &     $10+10i$ &  -2.49E+41 &  -3.30E+26 &  -2.49E+41 &   0.00E+00 &   5.00E+27 &   2.01E-14 \\

    21 &     $10+10i$ &  -5.23E+43 &  -5.23E+43 &  -5.23E+43 &  -5.23E+43 &   2.69E+30 &   3.63E-14 \\

    22 &     $10+10i$ &  -3.05E+32 &  -2.30E+46 &   0.00E+00 &  -2.30E+46 &   1.43E+33 &   6.22E-14 \\

    23 &     $10+10i$ &   5.29E+48 &  -5.29E+48 &   5.29E+48 &  -5.29E+48 &   1.17E+36 &   1.57E-13 \\

    24 &     $10+10i$ &   2.54E+51 &   1.70E+38 &   2.54E+51 &   0.00E+00 &   2.91E+39 &   1.15E-12 \\

    25 &     $10+10i$ &   6.35E+53 &   6.35E+53 &   6.35E+53 &   6.35E+53 &   2.15E+42 &   2.40E-12 \\

    26 &     $10+10i$ &   6.58E+42 &   3.30E+56 &   0.00E+00 &   3.30E+56 &   3.67E+44 &   1.11E-12 \\

    27 &     $10+10i$ &  -8.92E+58 &   8.92E+58 &  -8.92E+58 &   8.92E+58 &   8.63E+47 &   6.84E-12 \\

    28 &     $10+10i$ &  -5.00E+61 &   1.32E+47 &  -5.00E+61 &   0.00E+00 &   6.91E+50 &   1.38E-11 \\

    29 &     $10+10i$ &  -1.45E+64 &  -1.45E+64 &  -1.45E+64 &  -1.45E+64 &   2.47E+54 &   1.21E-10 \\

    30 &     $10+10i$ &   1.37E+53 &  -8.69E+66 &   0.00E+00 &  -8.69E+66 &   5.48E+56 &   6.31E-11 \\

    31 &     $10+10i$ &   2.69E+69 &  -2.69E+69 &   2.69E+69 &  -2.69E+69 &   5.89E+59 &   1.55E-10 \\

    32 &     $10+10i$ &   1.72E+72 &   0.00E+00 &   1.72E+72 &   0.00E+00 &   8.57E+62 &   4.97E-10 \\

    33 &     $10+10i$ &   5.69E+74 &   5.69E+74 &   5.69E+74 &   5.69E+74 &   4.84E+65 &   6.01E-10 \\

    34 &     $10+10i$ &  -1.65E+65 &   3.87E+77 &   0.00E+00 &   3.87E+77 &   1.64E+67 &   4.23E-11 \\

    35 &     $10+10i$ &  -1.35E+80 &   1.35E+80 &  -1.35E+80 &   1.35E+80 &   1.36E+75 &   7.11E-06 \\

    36 &     $10+10i$ &  -9.75E+82 &  -2.95E+69 &  -9.75E+82 &   0.00E+00 &   1.37E+74 &   1.40E-09 \\
    \hline
  \end{tabular}
\end{table*}

\begin{table*}
  \caption{\footnotesize{\textbf{Errors of Ryser's algorithm for real matrices.}
      Some data in FIG.~\ref{fig:errorsAllOneMatrices} comes from this table. We did the precision test by computing the permanent of an $n\times n$ all-$r$ real matrix, in which each element has a real value of $r$. In this case, the permanent in theory is $r^n\cdot n!$. In the table, $AbsErr=\left|Per_\text{Ryser}-Per_\text{Theory}\right|$ and $RelErr=\left|(Per_\text{Ryser}-Per_\text{Theory})/Per_\text{Theory}\right|$.
    }}
  \centering
  \label{tab:ERRORRyser}
  \begin{tabular}{c|ccccc|ccccc}
    \hline
    $n$ & $r$ & $Per_\text{Ryser}$ & $Per_\text{Theory}$ & $AbsErr$ & $RelErr$ & $r$ & $Per_\text{Ryser}$ & $Per_\text{Theory}$ & $AbsErr$ & $RelErr$ \\
    \hline
    17    & 1     & 3.56E+14 & 3.56E+14 & 3.08E+06 & 8.67E-09 & 0.1   & 3.56E-03 & 3.56E-03 & 1.26E-07 & 3.53E-05 \\
    18    & 1     & 6.40E+15 & 6.40E+15 & 7.86E+07 & 1.23E-08 & 0.1   & 6.40E-03 & 6.40E-03 & 3.74E-07 & 5.84E-05 \\
    19    & 1     & 1.22E+17 & 1.22E+17 & 8.14E+09 & 6.69E-08 & 0.1   & 1.22E-02 & 1.22E-02 & 4.90E-07 & 4.03E-05 \\
    20    & 1     & 2.43E+18 & 2.43E+18 & 3.80E+11 & 1.56E-07 & 0.1   & 2.43E-02 & 2.43E-02 & 2.01E-08 & 8.25E-07 \\
    21    & 1     & 5.11E+19 & 5.11E+19 & 2.94E+13 & 5.75E-07 & 0.1   & 5.11E-02 & 5.11E-02 & 5.78E-08 & 1.13E-06 \\
    22    & 1     & 1.12E+21 & 1.12E+21 & 2.08E+15 & 1.85E-06 & 0.1   & 1.12E-01 & 1.12E-01 & 1.93E-06 & 1.71E-05 \\
    23    & 1     & 2.59E+22 & 2.59E+22 & 5.99E+16 & 2.32E-06 & 0.1   & 2.59E-01 & 2.59E-01 & 1.22E-05 & 4.71E-05 \\
    24    & 1     & 6.20E+23 & 6.20E+23 & 1.15E+18 & 1.85E-06 & 0.1   & 6.20E-01 & 6.20E-01 & 4.60E-06 & 7.41E-06 \\
    25    & 1     & 1.55E+25 & 1.55E+25 & 2.27E+21 & 1.47E-04 & 0.1   & 1.55E+00 & 1.55E+00 & 7.27E-04 & 4.69E-04 \\
    26    & 1     & 4.03E+26 & 4.03E+26 & 4.62E+22 & 1.15E-04 & 0.1   & 4.03E+00 & 4.03E+00 & 5.97E-03 & 1.48E-03 \\
    27    & 1     & 1.09E+28 & 1.09E+28 & 2.51E+24 & 2.30E-04 & 0.1   & 1.11E+01 & 1.09E+01 & 1.75E-01 & 1.60E-02 \\
    28    & 1     & 3.05E+29 & 3.05E+29 & 3.26E+25 & 1.07E-04 & 0.1   & 2.93E+01 & 3.05E+01 & 1.14E+00 & 3.75E-02 \\
    29    & 1     & 8.67E+30 & 8.84E+30 & 1.73E+29 & 1.95E-02 & 0.1   & 7.57E+01 & 8.84E+01 & 1.27E+01 & 1.44E-01 \\
    30    & 1     & 2.85E+32 & 2.65E+32 & 1.94E+31 & 7.32E-02 & 0.1   & 6.15E+02 & 2.65E+02 & 3.50E+02 & 1.32E+00 \\
    31    & 1     & 7.58E+33 & 8.22E+33 & 6.39E+32 & 7.78E-02 & 0.1   & -4.19E+03 & 8.22E+02 & 5.01E+03 & 6.09E+00 \\
    32    & 1     & -2.49E+33 & 2.63E+35 & 2.66E+35 & 1.01E+00 & 0.1   & 6.65E+04 & 2.63E+03 & 6.39E+04 & 2.43E+01 \\
    33    & 1     & 8.35E+36 & 8.68E+36 & 3.31E+35 & 3.82E-02 & 0.1   & -4.16E+05 & 8.68E+03 & 4.25E+05 & 4.89E+01 \\
    34    & 1     & 3.60E+38 & 2.95E+38 & 6.44E+37 & 2.18E-01 & 0.1   & 2.20E+06 & 2.95E+04 & 2.17E+06 & 7.34E+01 \\
    35    & 1     & -5.42E+41 & 1.03E+40 & 5.52E+41 & 5.34E+01 & 0.1   & -6.35E+06 & 1.03E+05 & 6.45E+06 & 6.25E+01 \\
    36    & 1     & 4.24E+43 & 3.72E+41 & 4.21E+43 & 1.13E+02 & 0.1   & -3.05E+08 & 3.72E+05 & 3.06E+08 & 8.22E+02 \\
    17    & 10    & 3.56E+31 & 3.56E+31 & 8.65E+22 & 2.43E-09 & 0.2   & 4.66E+02 & 4.66E+02 & 2.75E-06 & 5.91E-09 \\
    18    & 10    & 6.40E+33 & 6.40E+33 & 1.95E+25 & 3.04E-09 & 0.2   & 1.68E+03 & 1.68E+03 & 1.29E-06 & 7.66E-10 \\
    19    & 10    & 1.22E+36 & 1.22E+36 & 4.69E+28 & 3.85E-08 & 0.2   & 6.38E+03 & 6.38E+03 & 2.64E-04 & 4.13E-08 \\
    20    & 10    & 2.43E+38 & 2.43E+38 & 7.63E+31 & 3.14E-07 & 0.2   & 2.55E+04 & 2.55E+04 & 8.12E-03 & 3.18E-07 \\
    21    & 10    & 5.11E+40 & 5.11E+40 & 3.07E+34 & 6.02E-07 & 0.2   & 1.07E+05 & 1.07E+05 & 4.65E-01 & 4.34E-06 \\
    22    & 10    & 1.12E+43 & 1.12E+43 & 2.19E+36 & 1.95E-07 & 0.2   & 4.71E+05 & 4.71E+05 & 8.81E+00 & 1.87E-05 \\
    23    & 10    & 2.59E+45 & 2.59E+45 & 2.73E+40 & 1.06E-05 & 0.2   & 2.17E+06 & 2.17E+06 & 1.06E+02 & 4.90E-05 \\
    24    & 10    & 6.20E+47 & 6.20E+47 & 3.83E+43 & 6.17E-05 & 0.2   & 1.04E+07 & 1.04E+07 & 8.29E+01 & 7.96E-06 \\
    25    & 10    & 1.55E+50 & 1.55E+50 & 8.36E+46 & 5.39E-04 & 0.2   & 5.21E+07 & 5.20E+07 & 2.44E+04 & 4.69E-04 \\
    26    & 10    & 4.03E+52 & 4.03E+52 & 6.24E+49 & 1.55E-03 & 0.2   & 2.70E+08 & 2.71E+08 & 4.00E+05 & 1.48E-03 \\
    27    & 10    & 1.10E+55 & 1.09E+55 & 8.37E+52 & 7.68E-03 & 0.2   & 1.48E+09 & 1.46E+09 & 2.34E+07 & 1.60E-02 \\
    28    & 10    & 2.95E+57 & 3.05E+57 & 9.56E+55 & 3.13E-02 & 0.2   & 7.88E+09 & 8.18E+09 & 3.07E+08 & 3.75E-02 \\
    29    & 10    & 9.49E+59 & 8.84E+59 & 6.46E+58 & 7.31E-02 & 0.2   & 4.06E+10 & 4.75E+10 & 6.84E+09 & 1.44E-01 \\
    30    & 10    & 1.92E+62 & 2.65E+62 & 7.31E+61 & 2.75E-01 & 0.2   & 6.61E+11 & 2.85E+11 & 3.76E+11 & 1.32E+00 \\
    31    & 10    & 1.07E+65 & 8.22E+64 & 2.50E+64 & 3.04E-01 & 0.2   & -9.00E+12 & 1.77E+12 & 1.08E+13 & 6.09E+00 \\
    32    & 10    & 6.54E+66 & 2.63E+67 & 1.98E+67 & 7.51E-01 & 0.2   & 2.86E+14 & 1.13E+13 & 2.74E+14 & 2.43E+01 \\
    33    & 10    & 3.60E+70 & 8.68E+69 & 2.73E+70 & 3.14E+00 & 0.2   & -3.58E+15 & 7.46E+13 & 3.65E+15 & 4.89E+01 \\
    34    & 10    & -8.76E+73 & 2.95E+72 & 9.06E+73 & 3.07E+01 & 0.2   & 3.77E+16 & 5.07E+14 & 3.72E+16 & 7.34E+01 \\
    35    & 10    & 1.32E+77 & 1.03E+75 & 1.31E+77 & 1.27E+02 & 0.2   & -2.18E+17 & 3.55E+15 & 2.22E+17 & 6.25E+01 \\
    36    & 10    & -2.14E+80 & 3.72E+77 & 2.14E+80 & 5.75E+02 & 0.2   & -2.10E+19 & 2.56E+16 & 2.10E+19 & 8.22E+02 \\
    \hline
  \end{tabular}
\end{table*}

\begin{table*}
  \caption{\footnotesize{\textbf{Errors of Ryser's algorithm for complex matrices.}
      We did the precision test by computing the permanent of an $n\times n$ all-$r$ complex matrix, in which each element has a complex value of $r$. In this case, the permanent in theory is $r^n\cdot n!$. $\text{Real}(Per_\text{Ryser})$ is the real part of $Per_\text{Ryser}$, and $\text{Imag}(Per_\text{Ryser})$ is the imaginary part. $AbsErr=\left|Per_\text{Ryser}-Per_\text{Theory}\right|$ and $RelErr=\left|(Per_\text{Ryser}-Per_\text{Theory})/Per_\text{Theory}\right|$.
    }}
  \centering
  \label{tab:ERRORRyserComplex}
  \begin{tabular}{cccccccc}
    \hline
    $n$ & $r$ & $\text{Real}(Per_\text{Ryser})$ & $\text{Imag}(Per_\text{Ryser})$ & $\text{Real}(Per_\text{Theory})$ & $\text{Imag}(Per_\text{Theory})$ & $AbsErr$ & $RelErr$ \\
    \hline
    17 &        $1+i$ &   9.11E+16 &   9.11E+16 &   9.11E+16 &   9.11E+16 &   1.12E+09 &   8.67E-09 \\

    18 &        $1+i$ &   0.00E+00 &   3.28E+18 &   0.00E+00 &   3.28E+18 &   4.03E+10 &   1.23E-08 \\

    19 &        $1+i$ &  -6.23E+19 &   6.23E+19 &  -6.23E+19 &   6.23E+19 &   5.89E+12 &   6.69E-08 \\

    20 &        $1+i$ &  -2.49E+21 &   0.00E+00 &  -2.49E+21 &   0.00E+00 &   3.89E+14 &   1.56E-07 \\

    21 &        $1+i$ &  -5.23E+22 &  -5.23E+22 &  -5.23E+22 &  -5.23E+22 &   4.25E+16 &   5.75E-07 \\

    22 &        $1+i$ &  -5.91E+17 &  -2.30E+24 &   0.00E+00 &  -2.30E+24 &   4.31E+18 &   1.87E-06 \\

    23 &        $1+i$ &   5.29E+25 &  -5.29E+25 &   5.29E+25 &  -5.29E+25 &   1.73E+20 &   2.32E-06 \\

    24 &        $1+i$ &   2.54E+27 &  -6.38E+21 &   2.54E+27 &   0.00E+00 &   7.92E+21 &   3.12E-06 \\

    25 &        $1+i$ &   6.35E+28 &   6.35E+28 &   6.35E+28 &   6.35E+28 &   1.33E+25 &   1.48E-04 \\

    26 &        $1+i$ &   1.44E+24 &   3.30E+30 &   0.00E+00 &   3.30E+30 &   6.83E+26 &   2.07E-04 \\

    27 &        $1+i$ &  -8.92E+31 &   8.92E+31 &  -8.92E+31 &   8.92E+31 &   2.90E+28 &   2.30E-04 \\

    28 &        $1+i$ &  -5.00E+33 &   2.40E+29 &  -5.00E+33 &   0.00E+00 &   5.85E+29 &   1.17E-04 \\

    29 &        $1+i$ &  -1.42E+35 &  -1.42E+35 &  -1.45E+35 &  -1.45E+35 &   4.00E+33 &   1.95E-02 \\

    30 &        $1+i$ &   3.91E+34 &  -9.33E+36 &   0.00E+00 &  -8.69E+36 &   6.37E+35 &   7.33E-02 \\

    31 &        $1+i$ &   2.48E+38 &  -2.48E+38 &   2.69E+38 &  -2.69E+38 &   2.96E+37 &   7.78E-02 \\

    32 &        $1+i$ &  -1.63E+38 &   0.00E+00 &   1.72E+40 &   0.00E+00 &   1.74E+40 &   1.01E+00 \\

    33 &        $1+i$ &   5.47E+41 &   5.47E+41 &   5.69E+41 &   5.69E+41 &   3.07E+40 &   3.82E-02 \\

    34 &        $1+i$ &   0.00E+00 &   4.71E+43 &   0.00E+00 &   3.87E+43 &   8.45E+42 &   2.18E-01 \\

    35 &        $1+i$ &   7.10E+46 &  -7.10E+46 &  -1.35E+45 &   1.35E+45 &   1.02E+47 &   5.34E+01 \\

    36 &        $1+i$ &  -1.11E+49 &   7.42E+47 &  -9.75E+46 &   0.00E+00 &   1.10E+49 &   1.13E+02 \\

    17 &     $10+10i$ &   9.11E+33 &   9.11E+33 &   9.11E+33 &   9.11E+33 &   3.13E+25 &   2.43E-09 \\

    18 &     $10+10i$ &   0.00E+00 &   3.28E+36 &   0.00E+00 &   3.28E+36 &   9.97E+27 &   3.04E-09 \\

    19 &     $10+10i$ &  -6.23E+38 &   6.23E+38 &  -6.23E+38 &   6.23E+38 &   3.39E+31 &   3.85E-08 \\

    20 &     $10+10i$ &  -2.49E+41 &   7.02E+33 &  -2.49E+41 &   0.00E+00 &   7.84E+34 &   3.15E-07 \\

    21 &     $10+10i$ &  -5.23E+43 &  -5.23E+43 &  -5.23E+43 &  -5.23E+43 &   3.98E+37 &   5.38E-07 \\

    22 &     $10+10i$ &  -2.36E+40 &  -2.30E+46 &   0.00E+00 &  -2.30E+46 &   2.72E+40 &   1.18E-06 \\

    23 &     $10+10i$ &   5.29E+48 &  -5.29E+48 &   5.29E+48 &  -5.29E+48 &   6.10E+43 &   8.15E-06 \\

    24 &     $10+10i$ &   2.54E+51 &  -4.61E+46 &   2.54E+51 &   0.00E+00 &   1.15E+47 &   4.54E-05 \\

    25 &     $10+10i$ &   6.36E+53 &   6.36E+53 &   6.35E+53 &   6.35E+53 &   4.44E+50 &   4.94E-04 \\

    26 &     $10+10i$ &   1.25E+52 &   3.30E+56 &   0.00E+00 &   3.30E+56 &   5.12E+53 &   1.55E-03 \\

    27 &     $10+10i$ &  -8.99E+58 &   8.99E+58 &  -8.92E+58 &   8.92E+58 &   9.69E+56 &   7.68E-03 \\

    28 &     $10+10i$ &  -4.84E+61 &   4.78E+58 &  -5.00E+61 &   0.00E+00 &   1.57E+60 &   3.14E-02 \\

    29 &     $10+10i$ &  -1.55E+64 &  -1.55E+64 &  -1.45E+64 &  -1.45E+64 &   1.50E+63 &   7.31E-02 \\

    30 &     $10+10i$ &   3.25E+64 &  -6.30E+66 &   0.00E+00 &  -8.69E+66 &   2.39E+66 &   2.76E-01 \\

    31 &     $10+10i$ &   3.51E+69 &  -3.51E+69 &   2.69E+69 &  -2.69E+69 &   1.16E+69 &   3.04E-01 \\

    32 &     $10+10i$ &   4.29E+71 &   0.00E+00 &   1.72E+72 &   0.00E+00 &   1.30E+72 &   7.51E-01 \\

    33 &     $10+10i$ &   2.36E+75 &   2.36E+75 &   5.69E+74 &   5.69E+74 &   2.53E+75 &   3.14E+00 \\

    34 &     $10+10i$ &  -5.63E+77 &   1.15E+79 &   0.00E+00 &   3.87E+77 &   1.11E+79 &   2.87E+01 \\

    35 &     $10+10i$ &  -1.73E+82 &   1.73E+82 &  -1.35E+80 &   1.35E+80 &   2.43E+82 &   1.27E+02 \\

    36 &     $10+10i$ &   5.60E+85 &   1.69E+82 &  -9.75E+82 &   0.00E+00 &   5.61E+85 &   5.75E+02 \\
    \hline
  \end{tabular}
\end{table*}

\begin{table*}
  \centering
  \caption{\footnotesize{\textbf{Errors of Ryser's algorithm from the order of add operations.}
      We did the precision test by computing the permanent of an $n\times n$ all-one real matrix, the permanent of which is $n!$ in theory. We tested different orders of add operations on intermediate results from different iterations. For example, $\sigma_0,\ldots \sigma_7$ represents eight intermediate results, $\prod_{i=1}^n\sum_{j \in S}a_{ij}$ in the iterations of Algorithm~\ref{alg:ParallelRyserProg}. Assume $\sigma_0,\sigma_2,\sigma_4,\sigma_8$ correspond to the cases of $(-1)^{|S|}=1$ in Algorithm~\ref{alg:ParallelRyserProg}, and $\sigma_1,\sigma_3,\sigma_5,\sigma_7$ correspond to that of $(-1)^{|S|}=-1$. ``Original'' performs $\sigma_0-\sigma_1+\sigma_2-\sigma_3+\sigma_4-\sigma_5+\sigma_6-\sigma_7$. ``Random'' performs add operations in a random order. ``Merging'' performs $(((\sigma_0-\sigma_1)+(\sigma_2-\sigma_3))+((\sigma_4-\sigma_5)+(\sigma_6-\sigma_7)))$. ``Separating'' performs $(\sigma_0+\sigma_2+\sigma_4+\sigma_6)-(\sigma_1+\sigma_3+\sigma_5+\sigma_7)$.
    }}
  \label{tab:ErrorOrder}
  \begin{tabular}{cccccccccc}
    \hline
    \multirow{2}{*}{$n$} & \multirow{2}{*}{$Per_\text{Theory}$} & \multicolumn{ 2}{c}{Original} & \multicolumn{ 2}{c}{Random} & \multicolumn{ 2}{c}{Merging} & \multicolumn{ 2}{c}{Separating} \\

    & & $Per_\text{Ryser}$ & $RelErr$ & $Per_\text{Ryser}$ & $RelErr$ & $Per_\text{Ryser}$ & $RelErr$ & $Per_\text{Ryser}$ & $RelErr$ \\
    \hline
    8 &   4.03E+04 &   4.03E+04 &   0.00E+00 &   4.03E+04 &   0.00E+00 &   4.03E+04 &   0.00E+00 &   4.03E+04 &   0.00E+00 \\

    9 &   3.63E+05 &   3.63E+05 &   0.00E+00 &   3.63E+05 &   0.00E+00 &   3.63E+05 &   0.00E+00 &   3.63E+05 &   0.00E+00 \\

    10 &   3.63E+06 &   3.63E+06 &   0.00E+00 &   3.63E+06 &   0.00E+00 &   3.63E+06 &   0.00E+00 &   3.63E+06 &   0.00E+00 \\

    11 &   3.99E+07 &   3.99E+07 &   0.00E+00 &   3.99E+07 &   0.00E+00 &   3.99E+07 &   0.00E+00 &   3.99E+07 &   0.00E+00 \\

    12 &   4.79E+08 &   4.79E+08 &   0.00E+00 &   4.79E+08 &   0.00E+00 &   4.79E+08 &   0.00E+00 &   4.79E+08 &   0.00E+00 \\

    13 &   6.23E+09 &   6.23E+09 &   0.00E+00 &   6.23E+09 &   0.00E+00 &   6.23E+09 &   0.00E+00 &   6.23E+09 &   0.00E+00 \\

    14 &   8.72E+10 &   8.72E+10 &   0.00E+00 &   8.72E+10 &   1.32E-07 &   8.72E+10 &   0.00E+00 &   8.72E+10 &   9.18E-09 \\

    15 &   1.31E+12 &   1.31E+12 &   1.96E-10 &   1.31E+12 &   3.62E-07 &   1.31E+12 &   4.86E-10 &   1.31E+12 &   2.35E-08 \\

    16 &   2.09E+13 &   2.09E+13 &   8.81E-10 &   2.09E+13 &   1.29E-07 &   2.09E+13 &   9.57E-10 &   2.09E+13 &   1.36E-07 \\

    17 &   3.56E+14 &   3.56E+14 &   1.11E-08 &   3.56E+14 &   9.41E-05 &   3.56E+14 &   8.24E-09 &   3.56E+14 &   4.81E-07 \\

    18 &   6.40E+15 &   6.40E+15 &   7.73E-09 &   6.40E+15 &   1.46E-04 &   6.40E+15 &   2.59E-09 &   6.40E+15 &   2.57E-06 \\

    19 &   1.22E+17 &   1.22E+17 &   4.28E-08 &   1.22E+17 &   1.72E-04 &   1.22E+17 &   8.13E-08 &   1.22E+17 &   7.31E-05 \\

    20 &   2.43E+18 &   2.43E+18 &   2.60E-07 &   2.39E+18 &   1.64E-02 &   2.43E+18 &   2.46E-07 &   2.43E+18 &   4.51E-04 \\

    21 &   5.11E+19 &   5.11E+19 &   9.52E-07 &   3.64E+19 &   2.88E-01 &   5.11E+19 &   1.01E-06 &   5.11E+19 &   9.39E-04 \\

    22 &   1.12E+21 &   1.12E+21 &   2.88E-06 &  -3.60E+20 &   1.32E+00 &   1.12E+21 &   4.28E-06 &   1.14E+21 &   1.59E-02 \\

    23 &   2.59E+22 &   2.59E+22 &   4.15E-06 &   6.96E+23 &   2.59E+01 &   2.59E+22 &   1.39E-05 &   2.72E+22 &   5.02E-02 \\

    24 &   6.20E+23 &   6.20E+23 &   1.59E-05 &  -1.21E+25 &   2.04E+01 &   6.20E+23 &   2.18E-05 &   7.80E+23 &   2.57E-01 \\

    25 &   1.55E+25 &   1.55E+25 &   1.05E-04 &  -5.49E+27 &   3.55E+02 &   1.55E+25 &   1.28E-04 &  -3.50E+25 &   3.26E+00 \\

    26 &   4.03E+26 &   4.03E+26 &   1.53E-04 &   2.29E+30 &   5.67E+03 &   4.03E+26 &   4.36E-04 &   1.68E+28 &   4.07E+01 \\

    27 &   1.09E+28 &   1.09E+28 &   5.48E-04 &  -1.36E+32 &   1.25E+04 &   1.09E+28 &   1.35E-03 &  -2.95E+30 &   2.72E+02 \\

    28 &   3.05E+29 &   3.05E+29 &   1.83E-04 &   1.61E+34 &   5.27E+04 &   3.05E+29 &   1.12E-04 &   4.66E+32 &   1.53E+03 \\

    29 &   8.84E+30 &   8.54E+30 &   3.47E-02 &  -6.42E+36 &   7.26E+05 &   8.47E+30 &   4.16E-02 &  -1.08E+35 &   1.22E+04 \\

    30 &   2.65E+32 &   2.97E+32 &   1.18E-01 &  -7.86E+38 &   2.96E+06 &   2.98E+32 &   1.23E-01 &   3.13E+37 &   1.18E+05 \\

    31 &   8.22E+33 &   6.15E+33 &   2.53E-01 &   7.00E+41 &   8.52E+07 &   5.81E+33 &   2.94E-01 &   1.75E+38 &   2.12E+04 \\

    32 &   2.63E+35 &   8.21E+34 &   6.88E-01 &  -8.38E+43 &   3.19E+08 &   1.24E+35 &   5.29E-01 &   2.65E+41 &   1.01E+06 \\
    \hline
  \end{tabular}
\end{table*}

\begin{table*}
  \centering
  \caption{\footnotesize{\textbf{Results comparing BB/FG's algorithm and Ryser's algorithm for $n\times n$ random unitary matrices.}
      Some data of ``Random unitary matrix'' in FIG.~\ref{fig:errorsRandomMatrix} comes from this table. $\text{Real}(Per)$ is the real part of $Per$, and $\text{Imag}(Per)$ is the imaginary part. $RelErr=\left|(Per_\text{Ryser}-Per_\text{BB/FG})/Per_\text{BB/FG}\right|$. The tested matrices are generated by MATLAB.
    }}
  \label{tab:ResultsComparing}
  \begin{tabular}{cccccc}
    \hline
    $n$ & $\text{Real}(Per_\text{Ryser})$ & $\text{Imag}(Per_\text{Ryser})$ & $\text{Real}(Per_\text{BB/FG})$ & $\text{Imag}(Per_\text{BB/FG})$ & $RelErr$ \\
    \hline
    31 & 4.2230E-09 & -1.9454E-09 & 4.2230E-09 & -1.9454E-09 &          0 \\

    32 & -4.1659E-09 & -2.3246E-09 & -4.1659E-09 & -2.3246E-09 &          0 \\

    33 & 4.8134E-10 & 8.4106E-10 & 4.8134E-10 & 8.4106E-10 &          0 \\

    34 & 2.7587E-10 & 2.1718E-10 & 2.7587E-10 & 2.1718E-10 &  2.848E-07 \\

    35 & -1.1053E-10 & -7.2113E-10 & -1.1053E-10 & -7.2113E-10 &  1.371E-07 \\

    36 & -5.5166E-11 & -3.4459E-11 & -5.5166E-11 & -3.4459E-11 &  4.612E-07 \\

    37 & -3.3585E-11 & 3.9724E-12 & -3.3585E-11 & 3.9722E-12 &  5.039E-06 \\

    38 & -5.6397E-11 & -6.3709E-11 & -5.6397E-11 & -6.3709E-11 &  2.351E-07 \\

    39 & -4.2777E-11 & -1.2135E-11 & -4.2777E-11 & -1.2135E-11 &  4.498E-07 \\

    40 & 5.8951E-12 & -2.1875E-11 & 5.8951E-12 & -2.1875E-11 &          0 \\
    \hline
  \end{tabular}
\end{table*}

\begin{table*}
  \centering
  \caption{\footnotesize{\textbf{Results comparing BB/FG's algorithm and Ryser's algorithm for random-derived matrices.}
      Some data of ``Random-derived matrix'' in FIG.~\ref{fig:errorsRandomMatrix} comes from this table. $\text{Real}(Per)$ is the real part of $Per$, and $\text{Imag}(Per)$ is the imaginary part. $RelErr=\left|(Per_\text{Ryser}-Per_\text{BB/FG})/Per_\text{BB/FG}\right|$. The tested matrices were derived from a $100\times 100$ unitary matrix.
    }}
  \label{tab:ResultsComparingDerived}
  \begin{tabular}{cccccc}
    \hline
    $n$ & $\text{Real}(Per_\text{Ryser})$ & $\text{Imag}(Per_\text{Ryser})$ & $\text{Real}(Per_\text{BB/FG})$ & $\text{Imag}(Per_\text{BB/FG})$ & $RelErr$ \\
    \hline
    21 & -4.3428E-12 & -5.9702E-12 & -4.3428E-12 & -5.9702E-12 &         0  \\

    22 & 7.6833E-13 & -3.9798E-13 & 7.6833E-13 & -3.9798E-13 & 1.1557E-07 \\

    23 & -3.7027E-12 & 1.6829E-12 & -3.7027E-12 & 1.6829E-12 &         0  \\

    24 & -3.4145E-12 & 1.6790E-12 & -3.4145E-12 & 1.6790E-12 &         0  \\

    25 & 6.3631E-13 & 6.0936E-13 & 6.3631E-13 & 6.0936E-13 &         0  \\

    26 & 2.0950E-13 & -3.6558E-13 & 2.0950E-13 & -3.6558E-13 &         0  \\

    27 & 1.7856E-13 & 1.7049E-13 & 1.7856E-13 & 1.7049E-13 &         0  \\

    28 & -1.4828E-14 & -5.6576E-14 & -1.4828E-14 & -5.6576E-14 &         0  \\

    29 & 9.6421E-14 & 2.0312E-13 & 9.6421E-14 & 2.0312E-13 &         0  \\

    30 & 1.7654E-14 & 2.7364E-14 & 1.7654E-14 & 2.7364E-14 &         0  \\

    31 & 2.7628E-16 & 1.4029E-14 & 2.7628E-16 & 1.4029E-14 & 7.1269E-09 \\

    32 & -1.7119E-14 & 1.3498E-14 & -1.7119E-14 & 1.3498E-14 &         0  \\

    33 & 6.9880E-15 & 4.0715E-15 & 6.9880E-15 & 4.0715E-15 &         0  \\

    34 & 1.0924E-15 & 1.2235E-14 & 1.0924E-15 & 1.2235E-14 &         0  \\

    35 & -1.0497E-16 & -1.3781E-16 & -1.0497E-16 & -1.3781E-16 &         0  \\

    36 & 5.9267E-16 & -1.9652E-16 & 5.9268E-16 & -1.9652E-16 & 3.2030E-06 \\

    37 & -3.5034E-16 & -2.3051E-16 & -3.5034E-16 & -2.3051E-16 &         0  \\

    38 & 2.7198E-15 & 6.0656E-16 & 2.7198E-15 & 6.0656E-16 &         0  \\

    39 & 7.9334E-16 & 5.6975E-16 & 7.9334E-16 & 5.6975E-16 & 2.4763E-06 \\

    40 & 1.2277E-16 & -2.4051E-16 & 1.2276E-16 & -2.4051E-16 & 2.0768E-05 \\

    41 & -1.2519E-16 & 1.7989E-16 & -1.2519E-16 & 1.7989E-16 & 1.2905E-06 \\

    42 & -5.1615E-16 & -3.3737E-16 & -5.1618E-16 & -3.3742E-16 & 9.6984E-05 \\

    43 & -8.8131E-17 & 3.3867E-16 & -9.0932E-17 & 3.3741E-16 & 8.7707E-03 \\

    44 & 1.3425E-18 & 1.0379E-19 & 1.3424E-18 & 1.0375E-19 & 5.9686E-05 \\
    \hline
  \end{tabular}
\end{table*}

\begin{table*}
  \centering
  \caption{\footnotesize{\textbf{Results comparing BB/FG's algorithm and Ryser's algorithm for special-derived matrices.}
      Some data of ``Special-derived matrix'' in FIG.~\ref{fig:errorsRandomMatrix} comes from this table. $RelErr=\left|(Per_\text{Ryser}-Per_\text{BB/FG})/Per_\text{BB/FG}\right|$. The tested matrices were manually chosen from the $100\times 100$ real unitary matrix with elements all negative, and duplicated the rows and columns to enlarge the scale.
    }}
  \label{tab:SpecialCase}
  \begin{tabular}{cccc}
    \hline
    $n$ & $Per_\text{Ryser}$ & $Per_\text{BB/FG}$ & $RelErr$ \\
    \hline
    30 & 5.3072E-01 & 4.6606E-01 & 1.3874E-01 \\

    31 & -2.2483E-01 & -1.2863E-01 & 7.4793E-01 \\

    32 & 1.1668E+00 & 7.4303E-01 & 5.7034E-01 \\

    33 & 6.3670E+01 & -3.6957E+00 & 1.8228E+01 \\

    34 & 3.5526E+01 & 2.1389E-01 & 1.6510E+02 \\

    35 & 1.4313E+04 & -7.9309E+02 & 1.9047E+01 \\

    36 & 2.8634E+04 & 3.1883E+02 & 8.8812E+01 \\

    37 & 6.3388E+04 & -7.0710E+01 & 8.9745E+02 \\

    38 & 1.3012E+07 & 2.5863E+03 & 5.0300E+03 \\

    39 & -1.0717E+07 & -4.8804E+03 & 2.1950E+03 \\

    40 & -1.0307E+08 & 1.2365E+03 & 8.3360E+04 \\
    \hline
  \end{tabular}
\end{table*}

\end{document}